\documentclass[english,twocolumn,a4paper,aps,showpacs,rmp]{revtex4-1}
\pdfoutput=1
\usepackage[T1]{fontenc}
\usepackage[latin1]{inputenc}
\usepackage[pdftex]{graphicx}
\usepackage{subfigure}
\usepackage{hyperref}
\usepackage{color}
\usepackage{amsfonts,latexsym}
\usepackage{amsmath,amssymb,bm}
\usepackage{babel}

\begin{document}  

\title{{\it Colloquium}: The transport properties of graphene: An introduction}

\author{N. M. R. Peres}

\affiliation{Physics Department and CFUM, University of Minho,  P-4710-057, Braga, Portugal}

\begin{abstract}
An introduction to the
transport properties of graphene combining
experimental results and theoretical analysis is presented. 
In the theoretical description  simple
intuitive models are used to illustrate important points on the transport
properties of graphene.
The concept of chirality, stemming
from the massless Dirac nature of the low energy physics of the
material, is shown to be instrumental
in understanding its transport properties:
the conductivity minimum, the electronic mobility, the effect of strain,
the weak (anti-)localization, and the
optical conductivity.
\end{abstract}



\maketitle

\tableofcontents{}


\section{Introduction}
\label{sec:introduction}

For a long time, theorists and
experimentalists alike have considered the existence of a true
two-dimensional (2D) material, having the
thickness of a single atom -- a one atom thick membrane -- to be impossible.
The reasoning behind this statement
relies on the fact that both finite temperature and quantum fluctuations
conspire to destroy the otherwise perfect 2D structure of the
hypothetic material. These fluctuations, originated from atomic
vibrations perpendicular to the  plane of the  material, would
preclude the existence of a true flat phase and concomitantly the existence
of such a system.

\begin{figure}[ht]
\includegraphics*[width=5.5cm]{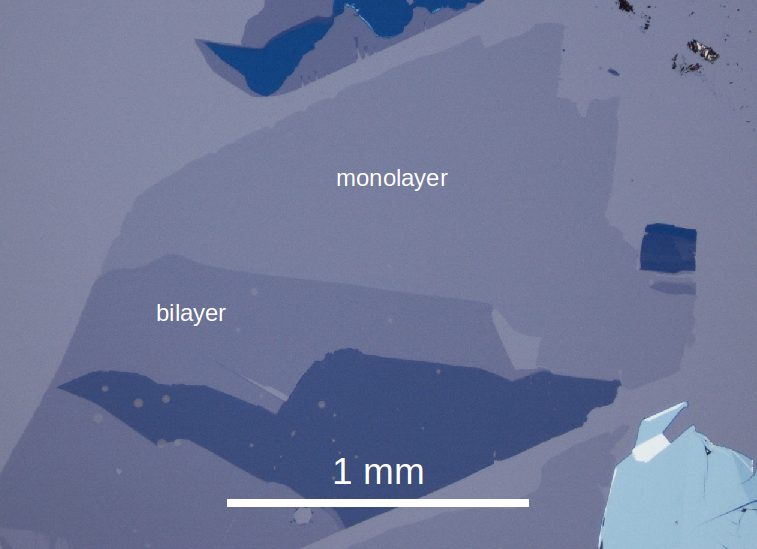}
\caption{(Color online)
An optical image of a graphene flake, obtained form the exfoliation of graphite,
with an area of $\lesssim 1$ mm$^2$,
 on top of a silicon oxide
wafer (courtesy of P. Blake).}
\label{fig:big}
\end{figure}

Nevertheless, in 2004, a group led by A. K. Geim,
from the University of Manchester, U.K., isolated such a 2D material
\cite{nov04,pnas}.
Under the name of graphene, this new
material is an allotropic form of carbon,
with the atoms arranged in a 2D honeycomb lattice. The reason for the success
lies on the isolation method. The developed method permitted one to isolate the 2D
material on top of
a 300 nm thick
wafer of silicon oxide.
The weak van der Waals interaction
induces adhesion between graphene and the wafer, and
once on top of the wafer, it is possible
to move about the 2D material, transferring it from one substrate to another, or
even
having it suspended over a trench, supported from one side \cite{geimstiffness}.
In the production method, graphite plays a key role, since this 3D material is
itself made of stacked  graphene planes (binded by van der Waals forces).
The ingenuity of the method
was then to find a way of peeling a single layer of graphene out
of graphite \cite{nov04,pnas}. Up to this date, the exfoliation
of graphite can produce graphene crystallites as large as $\sim$1 mm$^2$ (see
Fig. \ref{fig:big}).
 The study of graphene became, since 2004, an active
field of research in condensed matter, which holds many promises
\cite{geimprosp,service,scamerican,physW,natMat,mattoday,phystoday,peresEPN,
netocarbon,MRS}.

Being the first truly 2D material, it is natural to ask how its properties
differ from those of more conventional systems, such as the 2D electron
gas in the inversion layer of an ordinary semiconductor.
The current efforts in graphene research have  focused
 on the interplay among
 elastic, thermal, chemical, and
electronic properties of the material, with a special emphasis
on charge and heat transport, and on optical properties.
The need for a deep understanding of the transport properties of graphene
is  obvious, since
the material is a potential candidate
for incorporating  the future generation of nanoelectronic and nanophotonic
devices \cite{geimliquid,Avouris,Avouris1,Avouris2,bettergate,reviewontransistors}.
Also in biophysics, graphene is finding new applications \cite{biographene,grapheneDNA,DNAnose}.
Additionally, and of no less importance,
graphene provides a realm for the emergence
of  new and exciting physics.

In the field of electronic applications,
faster electronics requires
smaller devices, in particular because at the nanoscale it may be possible
for the electrons to travel across some of the components of a device
almost unimpeded.
In a normal conductor, one of  the sources of
electrical resistance is
scattering of electrons by impurities and defects
(and at room temperature, also by phonons).
A measure of the effect of impurities on the electronic
transport is the mean free path $\ell$
(the average distance traveled by an electron between two consecutive
collisions), which in a material with high degree of
purity and with small dimensions can be larger than the
typical length of the system $L_x$
leading, in these circumstances, to what is called ballistic
transport (in this regime the current becomes spatially non-uniform).
It just happens that in graphene  $\ell$ can be as large as 1 $\mu$m
\cite{nov04,kimsuspended},
putting graphene
into to the ballistic regime, since the typical size of graphene-based
field effect transistors is $L_x\sim$0.25-0.5 $\mu$m \cite{evasuspended}.

The first ground breaking  publications
 of the Manchester's group \cite{nov04,pnas} not only made
the method of isolating graphene  immediately public, but also established
the major relevant problems in graphene transport: the ambipolar field effect
(see Fig. \ref{fig:sigmacombined}), the
independence of the electronic mobility upon the gate voltage, the large
electronic
mean free path, the conductivity minimum and the absence of
Anderson localization \cite{bardarson}, the magneto-resistance and the
chiral quantum Hall effect \cite{qhegeim,qhekim}. These topics still orient
much of the research in graphene physics at  present.

Since the publication of a  comprehensive review on the
theoretical properties of graphene \cite{rmp}, there has been additional
relevant
 contributions to  experimental and theoretical studies of its transport
properties.
In this Colloquium, we present an update on the experimental
and theoretical developments in this fast growing subfield of graphene research,
at
a level appropriate to graduate students entering the field.

\section{The energy spectrum of graphene and the emergence of
Dirac electrons}
\label{sec:dispersion}

\begin{figure}[ht]
\includegraphics*[width=8cm]{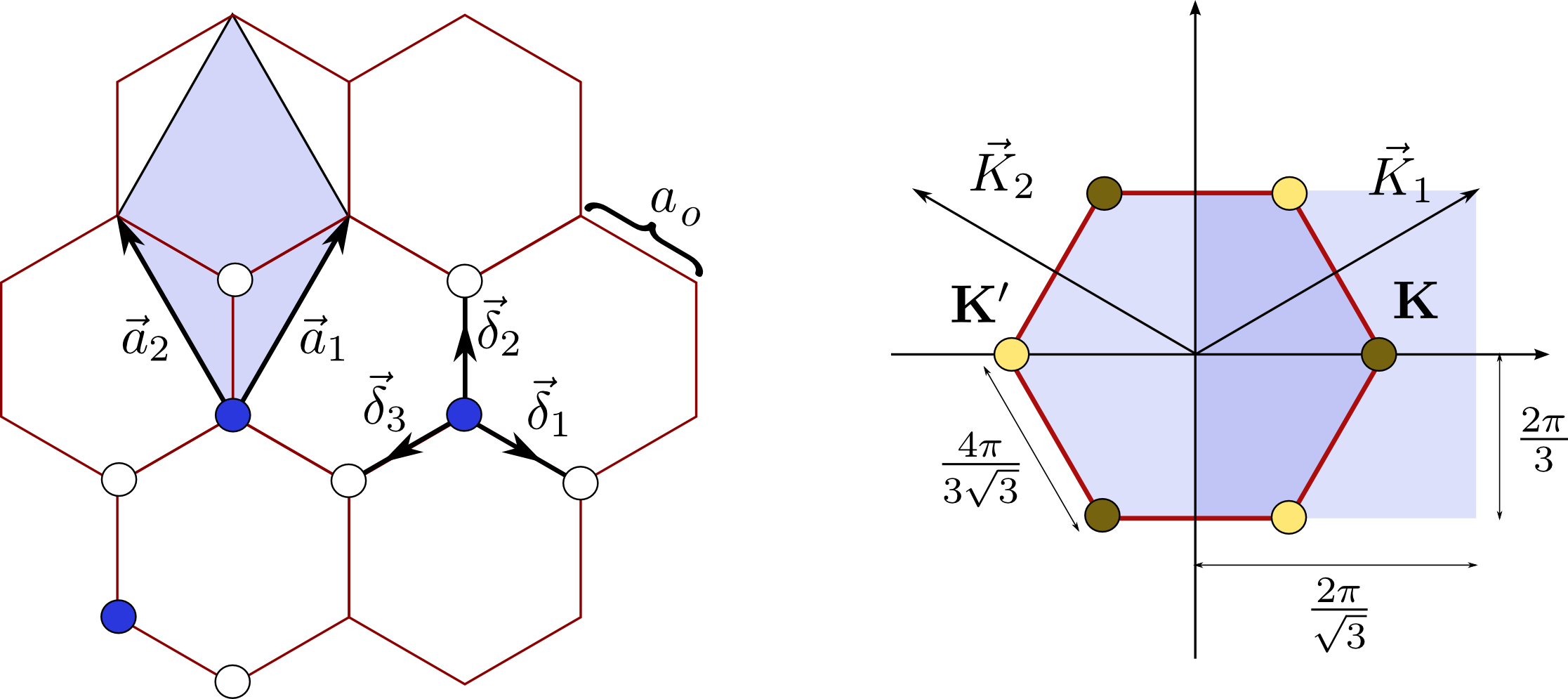}
\caption{(Color online)
Real space lattice and Brillouin zone of graphene.
{\bf Left:} The hexagonal lattice of graphene, with the nearest neighbor
$\bm \delta_i$ and the primitive  $\bm a_i$ vectors depicted.
The area of the primitive cell
is $A_c=3\sqrt{3}a_0^2/2\simeq 5.1$ \AA$^2$, and $a_0\simeq 1.4$ \AA.
{\bf Right:} The Brillouin zone of graphene, with the Dirac points
$\bm K$ and $\bm K'$ indicated. Close to these points, the dispersion
of graphene is conical and the density of states is proportional to the absolute
value
of the energy.}
\label{fig:lattice}
\end{figure}

As stated, graphene is a 2D material made solely of carbon atoms, arranged
in a hexagonal lattice such as that shown in Fig. \ref{fig:lattice}.
There are five vectors represented in Fig. \ref{fig:lattice}: the
three next-nearest neighbors vectors $\bm \delta_i$ ($i=1,2,3$), and the
primitive cell vectors $\bm a_1$ and $\bm a_2$. We further note
that the hexagonal lattice is made of two inter-penetrating triangular Bravais
lattices. Therefore, the effective model
describing the low-energy physics
of graphene has to keep track of the two atoms per unit cell, characteristic
of the honeycomb lattice.
Electrons in graphene can be described by a tight-binding
Hamiltonian reading (spin index omitted)
\begin{equation}
H=-t\sum_{n,\bm \delta_i}\vert A,\bm R_n\rangle \langle \bm R_n+\bm\delta_i,
B\vert
+{\rm H.\,c}.\,,
\label{eq:tbhamilt}
\end{equation}
where $\vert \bm A,\bm R_n\rangle$ represents the Wannier state
at the unit cell $\bm R_n$, and the equivalent definition
holds for $\vert  B,\bm R_n+\bm \delta_i\rangle$; $t$ is the hopping energy.
This Hamiltonian  describes the
motion of  electrons in the
$\pi-$orbitals of the material,
made from the hybridization of the atomic 2$p_z$ orbitals,
and includes both low-energy and high-energy electron
states.
The   calculation of the electronic energy spectrum of graphene
 proceeds by introducing, in Eq. (\ref{eq:tbhamilt}), the
Fourier representation of the Wannier states in terms of the Bloch states of
momentum $\bm k$; the spectrum then reads \cite{Wallace:1947,rmp}
\begin{equation}
E(\bm k)=
\pm t\vert 1+ e^{i\bm k \cdot \bm a_1} + e^{i\bm k \cdot \bm a_2}\vert\,.
\label{eq:spectrum}
\end{equation}
It is immediately obvious that the band structure of the $\pi-$electrons
is composed
of two bands, one at negative energies (a hole  band)
and the other at positive ones (a particle  band).
In the Brillouin zone there are two special, non-equivalent (i.e.
not connected by  a reciprocal lattice vector),
 wave numbers,
termed $\bm K$ and $\bm K'$, and shown in Fig. \ref{fig:lattice}.
The transport properties of graphene are mostly determined by
the nature of the spectrum around these two points.
Close to  $\bm K$ and $\bm K'$
the dispersion, Eq. (\ref{eq:spectrum}), is
conical, and given by
 $E(\bm k)=\pm v_F\hbar k$,
with $v_F=3ta_0/2\hbar$, where
$k$ is the momentum measured relatively to either
$\bm K$ or $\bm K'$, depending on the position of the cone in the Brillouin
zone.
Using the widely accepted value of $t\simeq -2.7$ eV for the hopping
(in reality the values of $t$ vary in the literature, spanning the interval
from -2.7 to -3.1 eV)  we obtain $v_F\lesssim 10^{6}$ m/s. The experimental
studies
are consistent in obtaining $v_F\simeq 1.1\times10^{6}$ m/s
\cite{qhegeim,qhekim,STMgiantphonon,Landauspectroscopy}.
A direct measurement of the
Dirac spectrum in graphene has recently been obtained using
angle-resolved photo-emission spectroscopy \cite{lanzara}.
Since each carbon atom (electronic configuration
1$s^2$ 2$s^2$ 2$p^2$) hybridizes with its three nearest neighbors
according to the hybrid orbitals  $sp^2$, there is one electron
left in the $p_z$ orbital. Therefore, the system is  half filled, with
the important consequence that
the low-energy physics is controlled by the spectrum close to the
$\bm K$ and $\bm K'$ points. Many of the new and exciting properties
of graphene stem from this fact.
The vicinities of these two points are also referred to
as the two valleys of the  electronic
spectrum of graphene.

The spectrum $E(\bm k)=\pm v_F\hbar k$ is formally equivalent to that
obtained from solving the 2D massless Dirac equation. Indeed, it is
easy to
show \cite{rmp,semenoff} that close to the $\bm K$  point the effective
Hamiltonian for the electrons in graphene has the form
\begin{equation}
 H_{\bm K}=v_F\bm \sigma\cdot\bm p\,,
\label{eq:dirac}
\end{equation}
whereas close to $\bm K'$, the Hamiltonian is obtained from Eq.
(\ref{eq:dirac})  by making the transformation $H_{\bm K'}=-H_{\bm K}$.
The operator $\bm \sigma$ is written in terms of the
Pauli matrices as
$\bm \sigma=(\sigma_x,\sigma_y)$, and $\bm p$ is the
momentum operator. Computing the eigenvalues of the Hamiltonian
(\ref{eq:dirac}),  the conical spectrum indicated above is immediately obtained.
We  stress that  $\bm \sigma$ does not
represent real electronic spin; it is instead a formal way of
taking into account the two carbon atoms per unit cell in graphene,
as we have anticipated above. For this
reason,  $\bm \sigma$ is termed pseudo-spin.
The density of states associated
with the conical dispersion  of  electrons in graphene is computed by
determining the number of states per unit cell in the Brillouin zone $N(E)$
up to the momentum $k$. Taking into account
contributions from states near  $\bm K$  and $\bm K'$ points, we obtain
 $N(E)= k^2A_c/(2\pi)$, from which the density of states $\rho(E)$ per spin
and per unit cell is given by
$
\rho(E)\equiv d\,N(E)/d\,E=2\vert E\vert/(\pi\sqrt{3}t^2)\,,
$
and
the primitive cell area, $A_c$, is defined in the caption of Fig. \ref{fig:lattice}.
The linear dependence of the density of states on energy is one of the
fingerprints
of massless Dirac electrons. For neutral graphene, the Fermi energy is zero.
Therefore,
the density of states vanishes in this case.

\begin{figure}[ht]
\includegraphics*[width=7cm]{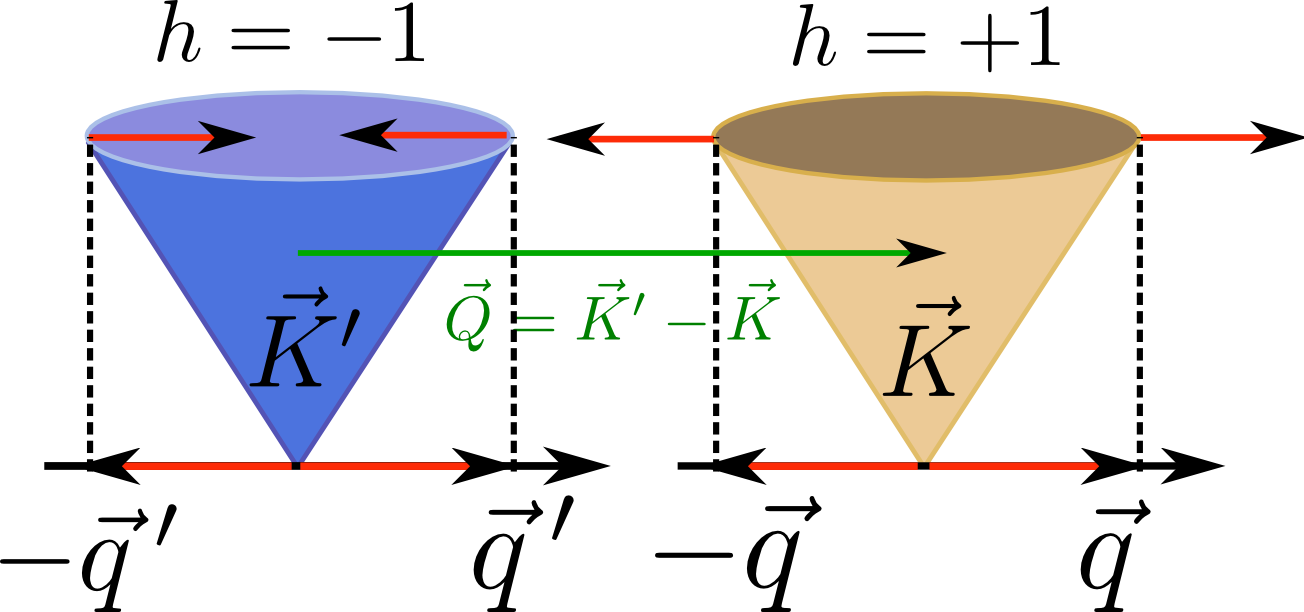}
\caption{(Color online)
At the $\bm K$ valley,  electrons have positive helicity, $h=1$,
whereas at the $\bm K'$ one, the helicity is negative ($\bm Q=\bm K'-\bm K$
represents the transferred momentum
when a scattering event between the valleys takes place).
In a head-on collision of the electron on a potential barrier, the backscattered
electron has to change its momentum from $\bm q$ to $-\bm q$.
For such a head-on collision (taken here along the 
$x-$direction), $\hat h$ 
is a constant of motion, with eigenvalue $+1$, but backscattering would imply 
a modification of this eigenvalue to $-1$. This, however, cannot be, because
$\hat h$ is a conserved quantity,  
then
the transmission probability through the barrier, for
such type of collision,
has to be one.
Thus, backscattering is suppressed for intra-valley scattering events.
On the other hand, electrons in the $\bm K'$ and $\bm K$ valleys have
opposite chirality,
thus inter-valley backscattering can take place (if the potential is short range), since
in this case  the eigenvalue of $\hat h$ does not change sign. This discussion
will be of importance
for Sec. \ref{sec:weakgraph}. 
}
\label{fig:scattering}
\end{figure}

The electronic linear spectrum and the chiral nature
of the electron's wave function (see below) make
electronic behavior in graphene quite unique, and are
responsible for the remarkable properties of this material.
Since  $\bm\sigma\cdot\bm p\vert\psi\rangle=\pm p\vert\psi\rangle$,
then the operator $\hat h=\bm\sigma\cdot\bm p/p$ has only two eigenvalues
$\pm 1$. The operator $\hat h$ is known as the helicity operator,
and has the following physical interpretation: in an energy eigenstate, the
pseudo-spin
 $\bm \sigma$ is either
parallel or anti-parallel to the momentum $\bm p$.
In the $\bm K$ valley, electrons have positive helicity and holes have negative
helicity; in $\bm K'$ the opposite happens.
The helicity (or chirality) of electrons in graphene
is responsible for the
Klein tunneling effect \cite{beenakkrmp,cheianovKlein,klein},
observed recently in graphene heterojunctions \cite{goldklein,kimklein}.
We then see (and at odds to high-energy
neutrino physics) that  massless Dirac electrons in graphene come
with both right and left chirality: parity is a symmetry of
graphene. In Fig. \ref{fig:scattering} we show, in simple terms, the
origin
of the Klein tunneling effect: the probability of electronic transmission
through a potential
barrier is equal to 1, for head-on collisions; it is said that backscattering
is suppressed.

We should note that chirality is not, however, an exact symmetry
of the problem. This occurs because the spectrum of graphene
is not exactly linear at all energies. The deviation
from the perfect massless Dirac behavior is known as
trigonal warping \cite{mccann,narozhny}, and
starts playing a role for energies $E\gtrsim1$ eV.
We  remark, however, that
trigonal warping might be
important for observation of weak
localization at energies much lower than 1 eV (see Sec. \ref{sec:weakgraph}).

The solution of the
eigenproblem $H_{\bm K}\vert\psi\rangle=E\vert\psi\rangle$ is easily obtained by
recognizing its
formal equivalence to that of a real spin in a magnetic field \cite{rmp}, with
the
wave function reading
\begin{equation}
 \vert\psi_\pm\rangle=\frac{1}{\sqrt{2}}\left(
\begin{array}{c}
e^{-i\theta(\bm k)/2}\\
\pm e^{i\theta(\bm k)/2}
\end{array}
\right)e^{i\bm k\cdot\bm r}\equiv u_\pm(\bm k)e^{i\bm k\cdot\bm r}\,,
\label{eq:spinora}
\end{equation}
and $\theta(\bm k)=\arctan(k_y/k_x)$.
Since the eigenproblem we have just solved is formally identical
to a spin one-half in a magnetic field, the spinors change
sign upon the transformation $\theta(\bm k)\rightarrow\theta(\bm k)+2\pi$,
as dictated by the spin-statistics theorem.

The first strinking consequence of the chiral nature of electrons in
graphene was the observation of the chiral quantum Hall effect
\cite{qhegeim,qhekim},
where the Hall conductivity is quantized as  $\sigma_{xy}=2e^2(1+2n)/h$,
with $n=1,2,\ldots$ \cite{nmrPRB06,gusynin3}. The quantization rule follows
from the nature of the Landau levels of Dirac electrons
\cite{rabi,lippmann,nieto,pereslandau}
combined with the existence of the two valleys in graphene.

The application of the chiral quantum Hall effect to metrology,
in defining the resistance standard \cite{metrologia},
has clear advantages over the usual quantum Hall effect in the 2D
electron gas, since the same experimental accuracy on the quantization of the
Hall
resistance  can be achieved  at higher temperatures
\cite{geimmetrology,metrologyNatMat,metrologyNatMatNewsandViews}.
At a temperature of 300 mK, the accuracy of the quantum Hall resistance
quantization  has been shown to be of
3 parts per billion, in monolayer epitaxial graphene
\cite{metrologyNatMat,metrologyNatMatNewsandViews}.
Also,  the
quantum Hall effect in graphene has been observed at room temperature
\cite{qheffectT}
and recently  in epitaxial graphene as well \cite{deheer}, which can be
produced in
quasi-free standing form \cite{epitaxialquasifree}.

Electron-electron interactions play no role in the
half-integer or chiral quantum Hall effect. On the
other hand, they are a crucial ingredient in the
explanation of the fractional quantum Hall effect.
 During the first few years of graphene research, effects of electron-electron
interactions have been elusive,
 but the recent observation of the 1/3 fractional Hall plateau
\cite{Evafrac,Kimfrac,morpurgoQHE},
brings them to the forefront
this  active research area. It is a remarkable experimental fact
that the fractional quantum Hall effect in graphene can be observed at magnetic
fields of 2 T
and persists up to a temperature of 20 K, for fields of 12 T.

Using the  results introduced above, we  proceed to the
discussion of several topics on electronic transport in graphene.

\section{ Conductivity and conductance of graphene at the Dirac point}
\label{sec:min_bulk}

As discussed in Sec. \ref{sec:dispersion},
undoped graphene has its Fermi energy at the Dirac point, where
the material has a vanishing density of states.
This would naively suggest that the
conductivity of undoped graphene should be zero.
However, experiments challenge ones intuition and show a finite conductivity
at zero energy (i. e., at the neutrality or Dirac point).
An example of a conductivity curve of graphene is shown in
 Fig. \ref{fig:chargescatterers}, where we see that the
experimental conductivity minimum, at $V_g=0$, is of the order of $\sim 4e^2/h$
(horizontal dashed line).
Values of the conductivity minimum for several
devices are given in Fig. \ref{fig:sigma_minimum}.
The existence of a conductivity minimum in graphene is also referred to as
quantum-limited resistivity.

\subsection{Sources of disorder}
\label{sec:sourcesdisorder}

As in any other metallic system, the electronic mobility in graphene is
hindered by disorder.
The sources of disorder in graphene can vary, and
can be due to
 adsorbed atoms (for example hydrogen)
or molecules (for example hydrocarbons),
extended
defects, such as folded regions (wrinkles), vacancies, and topological defects
[such as of Stone-Wales type, specially at the edges \cite{stonewales}].
Interestingly enough,
in some particular cases, an extended defect in graphene can act as a 1D
conducting channel
\cite{metalicwire}.
In addition, the system has a certain amount of rippling (random strain)
\cite{meyerGeimsuspended,kastripples}, so it is not
a perfect planar lattice, and
it has rough edges, which can exhibit scrolling \cite{foglerscroll}.
We should note that,
although the formation of vacancies is energetically unfavorable, the existence
of adatoms and adsorbed hydrocarbons is likely, originating from the
isolation method and exposure to the environment.
Such adsorbed atoms can be imaged by transmission electron
microscopy \cite{meyerdisorder}.
Additionally,
the electrostatic random potential at the surface of the
silicon-oxide substrate  acts as an additional scattering source, originated
from
 charged impurities \cite{Cromieinhomogeneous}.

To a good practical approximation, an
 adsorbed hydrocarbon, when binding covalently to the 2$p_z$ orbital
of a given carbon atom of graphene,
effectively removes the  2$p_z$ electron from participating
in the electric transport, by forming a $\sigma-$bond.
Since the electron wave-function is spatially confined, the impurity
can effectively act as a vacancy.
This latter type of defects induce
resonant states at, or close to, the Dirac point (see  below).

Another way of looking at this problem is to consider that, say, an
hydrogen atom  when binding covalently \cite{Ishigami} to a carbon atom in graphene
changes locally the hybridization from pure $sp_2$ to partially $sp_3$ and
creates, as before,
a resonant impurity at that site \cite{robinson,netospinorbit}.
In this latter sense, both
local potentials and adatoms have a similar effect \cite{stauberphonons}.
The change of the chemical bonds from pure $sp_2$ to partially $sp_3$ 
adds an additional scattering effect originated from  the enhancement of spin-orbit
coupling \cite{netospinorbit}.

Combined with charged scatterers, the resonant scattering mechanism
is currently ascending as one of the dominant processes limiting
the electronic mobility in graphene \cite{Dpeak}.

The resonant scattering mechanism
is easy to understand by considering a simple model. We add to
the Hamiltonian (\ref{eq:tbhamilt}) a contribution from an impurity binding
covalently to a carbon atom at site $\bm R_n=0$. Such a situation
adds to the Hamiltonian a term of the form
$H_{\rm rs}=(V\vert {\rm ad}\rangle\langle A,0 \vert+{\rm H.\,c.})
+\epsilon_{\rm ad}\vert {\rm ad}\rangle\langle {\rm ad} \vert$,
where $V$ is the hybridization between
the adatom (or a carbon atom of a hydrocarbon molecule)
and a given carbon atom of graphene, $\epsilon_{\rm ad}$
is the relative (to graphene's carbon atoms)
on-site energy of the electron in the
adatom, and $\vert {\rm ad}\rangle$ is the ket representing the state of the
electron in the adatom.
Taking the wave function to be of the form
$\vert\psi\rangle=\sum_{n}[A(\bm R_n)\vert A,\bm R_n \rangle+
B(\bm R_n+\bm\delta_2)\vert B,\bm R_n +\bm\delta_2\rangle]+
C_{\rm ad}\vert {\rm ad}\rangle$, the Schr\"odinger
equation at the site $\bm R_n=0$ reads
\begin{eqnarray}
EA(0)-VC_{\rm ad}&=&-t[B(\bm\delta_1)+B(\bm\delta_2)+B(\bm\delta_3)]\,,\\
(E-\epsilon_{\rm ad})C_{\rm ad}&=&VA(0)\,.
\end{eqnarray}
Solving for $C_{\rm ad}$, we obtain
\begin{equation}
-t[B(\bm\delta_1)+B(\bm\delta_2)+B(\bm\delta_3)] =
EA(0)-\frac{V^2A(0)}{E-\epsilon_{\rm ad}}\,.
\label{eq:resonantTB}
\end{equation}
The resonant effect is included in the last term of Eq. (\ref{eq:resonantTB}),
which
represents a local potential of the form $V_{\rm eff}=V^2/(E-\epsilon_{\rm
ad})$.
Equation (\ref{eq:resonantTB}) contains two interesting regimes: (i) when
$\vert E\vert \ll \epsilon_{\rm ad}$,
the adatom acts as an effective local potential of strength
$g_{\rm eff}=V^2/\epsilon_{\rm ad}$. If $g_{\rm eff}$ is large, the adatom
acts roughly as
an effective vacancy; a vacancy is characterized by  $g_{\rm eff}=\infty$; (ii)
when
$E\approx\epsilon_{\rm ad}$, the hopping from the  carbon atom at position $\bm
R_n=0$
to its nearest neighbors is
suppressed [effectively we have $t\rightarrow (E-\epsilon_{\rm ad})t$],
and the adatom acts roughly and again as
an effective vacancy at energies close to $\epsilon_{\rm ad}$.
Therefore, either by inducing an effective local potential or by suppressing
the nearby hopping we see that such mechanism increases the likelihood
of an electron being trapped for a longer time in the vicinity of the adatom,
thus
generating a resonant state.

If $\epsilon_{\rm ad}\simeq 0$, then the resonant states will be
exactly at the Dirac point, and this is expected to happen for adsorbed
hydrocarbon molecules.
It is then  the job of  quantum chemical calculations
to determine the value of the
parameters $\epsilon_{\rm ad}$ and $V$ \cite{robinson,wehling,wehlingII}. 
Recently obtained
 typical values are $V\sim 2t\sim 5$ eV and $\epsilon_{\rm ad}\sim$ -0.2
\cite{wehlingII},
leading to $g_{\rm eff}\sim 100$ eV, a rather strong on-site potential.
Finally,
the calculation of the transport properties for such a model can be performed
using the
$T-$matrix approach \cite{tsai1,tsai2,robinson}. Its derivation is elementary,
using the simple
 model described above. It is well known that the $T$ matrix for a local
potential of intensity
$v_0$ reads \cite{nmrPRB06,bena}
$
T(E)=v_0[1-v_0\bar G_R(E)]^{-1}$.
Then, for an adatom we must have
\begin{equation}
T(E)=\frac{V_{\rm eff}}{1-V_{\rm eff}\bar G_R(E)}=\frac{V^2}{E-\epsilon_{\rm
ad}-V^2\bar G_R(E)}\,.
\label{eq:rstmatrix}
\end{equation}
Using Eq. (\ref{eq:rstmatrix}), it is simple to compute the transport relaxation
time $\tau(\epsilon_F)$
 at the Fermi energy $\epsilon_F$ using
$
\hbar/\tau(\epsilon_F) = \pi n_i\vert T(\epsilon_F)\vert^2\rho(\epsilon_F)\,,
$
where $n_i$ is the concentration of impurities per unit cell.
From the knowledge of $\tau(\epsilon_F)$, the conductivity of graphene follows
from Boltzmann's transport equation \cite{ziman}
(see Sec. \ref{sec:sigma_gated}). The function $\bar G_R(E)$ reads:
$\bar G_R(E)=ED^{-2}
\ln (E^2/D^2)
-i\pi \vert E\vert/D^2$, with $D\simeq 3t$.

It has been theoretically predicted that,
in addition to their scattering effect, monovalent adatoms in diluted
concentrations
can
create a gap in graphene's spectrum, by a
mechanism called {\it sublattice ordering} \cite{cheianovRS}.
Superlattices of vacancies (or adatoms) have the same effect \cite{Simonevacancygap}.

  Midgap states \cite{jackiw}
are also produced by a model of pure vacancies \cite{vitordisorder,vitorpaco},
as shown in Fig. \ref{fig:CPA},
and if a nearest neighbor hopping ($t'\simeq$0.4 eV) is included, the resonant
states, while no longer at exactly the Dirac point, remain at energies close
to it \cite{vitordisorder,vitorpaco}.

\subsection{Calculation of the conductivity minimum for bulk graphene
due to disorder}
\label{sec:disorder}

It is certainly
 difficult to model all the different types of disorder just mentioned in a
single
calculation.
We for the moment ignore this complexity and assume
 that electrons in graphene move in a random potential of the form
$
V(\bm r)=v_0\sum_{n=1}^{N_i}\delta(\bm R_n-\bm r)\,,
$
 where the position vectors $\bm R_n$ are random, $v_0$ is the strength of the
potential, and $N_i$ is the number of scattering centers.
This model can be seen in the worst case scenario as zero order description of
the effect of impurities in graphene, although
it has  recently been  used widely
\cite{andoberry2,andoWL,andozheng,nmrPRB06,OstrovskySCBA,OstrovskyEJP}.
In fact, for large $v_0$ this model mimics the resonant scatterers physics.
In what follows, we  determine the consequences of the above  random
potential  on the minimum conductivity of
graphene.

\begin{figure}[th]
\includegraphics*[width=8cm,angle=0]{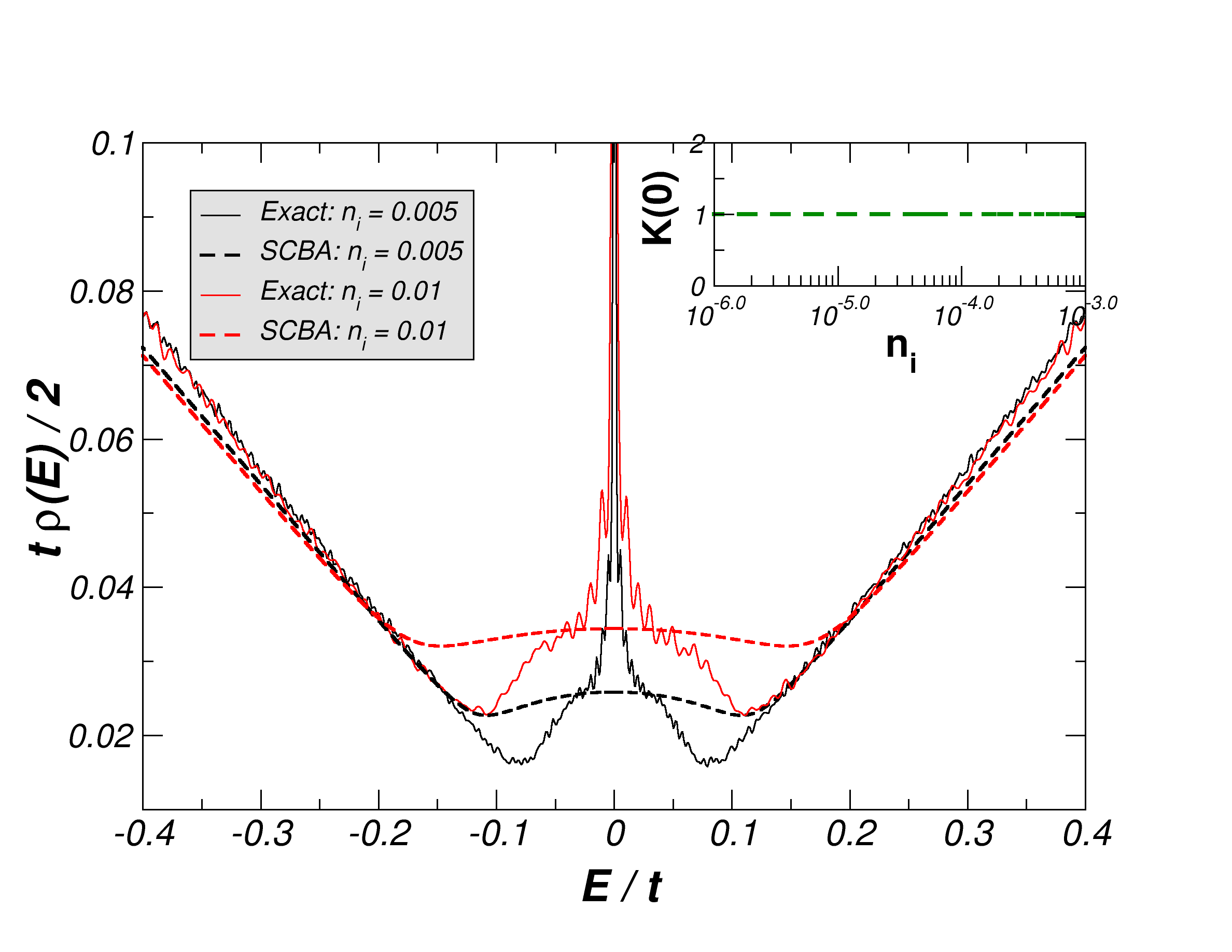}
\caption{(Color online)
Numerically exact density of states (solid lines),
in the limit $v_0\rightarrow\infty$, and  the corresponding
SCBA calculation
(dashed lines), for the same impurity concentrations $n_i$.
The maximum of the SCBA density of states, at $E=0$,
follows the rule
$\rho(E)\simeq 0.2\sqrt{n_i}$ eV$^{-1}$.
{\bf Inset:}
 The function $K(0)$ is plotted for a  range
of impurity concentrations spanning three orders of magnitude.
(The  numerically exact
calculations are courtesy of Vitor M. Pereira.)}
\label{fig:CPA}
\end{figure}

The usual approach to the calculation of the conductivity uses the
Kubo-Greenwood formula, obtained from linear response theory
\cite{mahan}. The calculation proceeds in two steps
\cite{nmrPRB06,andoshon,andozheng}:
first,
the single particle Green's function in the presence
of the disordered potential is computed in a self-consistent
manner;
second the current-current correlation function is obtained in terms
of the single particle Green's function. This method is known as the
self-consistent Born
approximation (SCBA).
The final result of such  calculation is a  simple expression
for the conductivity $\sigma(\epsilon_F)$ at the Fermi energy reading
\begin{equation}
\sigma(\epsilon_F)=\frac{4e^2}{\pi h}K(\epsilon_F)\,,
\label{eq:sigmamin}
\end{equation}
where $K(\epsilon_F)$ is a dimensionless function
\cite{nmrPRB06,stauberphonons}; Eq. (\ref{eq:sigmamin}) holds
true both at finite $v_0$ or when $v_0\rightarrow\infty$; from here on
we consider this latter regime only. Since we  describe the
transport at the Dirac point, we need the value of $K(\epsilon_F)$
at zero chemical potential, which turns out to be $K(0)\simeq 1$.
This result is essentially insensitive to the concentration
of impurities $n_i=N_iA_c/A$ ($A$ is the area of the sample
and $n_i$ is the concentration of impurities per unit cell).
The behavior of $K(0)$ as function of $n_i$ is shown in the
inset of Fig. \ref{fig:CPA}; as stated, its value is 1.
We have, therefore, obtained a universal value for the conductivity
minimum of graphene
$\sigma_{\rm min}=4e^2/(\pi h)$ independent of
the impurity concentration, even if the concentration of impurities is
a  small number. Many  have reached the same result using different
approaches \cite{klausmin,klausdos}. A question naturally arises: How
does one understand the result given by Eq. (\ref{eq:sigmamin})? To that end, we
compute the density of states of disordered graphene.

We have compared a calculation of the density of states as given by
the SCBA with that given by an exact numerical method
\cite{vitordisorder,vitorpaco}.
In Fig. \ref{fig:CPA}
we show two sets  of calculations for the density of states close to the Dirac
point
($E\sim$0).  Two features are
clear from these calculations. First, the disorder only
affects the DOS close to the Dirac point, rendering it finite; second
the SCBA introduces a smoothing of  DOS around $E\sim$0, but
its value essentially agrees with the exact one, except at energies
very close to $E= 0$. The finite density of states close to the Dirac point
is due to the wings of the resonant states forming at zero energy.
The same behavior is seen in the local density of states around a single vacancy
and in the corresponding scanning tunnelling microscopy current
\cite{tsai1,tsai2}.

The above comparison  shows that the SCBA gives a reasonable description
of the density of states close  to the Dirac point, and this gives us a certain
amount
of confidence in the calculation of the conductivity $\sigma(\epsilon_F)$
based on the same approximation. A comment on the behavior of the
numerical DOS close to zero energy is in order:
the sharp feature at precisely $E=0$ seen in the exact numerical solution
arises from the presence of zero-energy quasilocalized modes,
induced by the vacancies in the lattice \cite{vitordisorder,vitorpaco}. These
localized states are clearly not captured by the SCBA.

In short, the finiteness of the conductivity at the Dirac
point is a consequence of the finiteness of the DOS at $E\sim$0
due to disorder, even when the concentration of impurities is small,
since $K(0)$ is essentially constant
over several orders of magnitude of  impurity concentration.
Furthermore,
there is a strong criticism  in the literature regarding the
application of the SCBA approach to describe the physics at $\epsilon_F\simeq 0$
\cite{aleiner}, but not at finite $\epsilon_F$, as long as weak localization
effects are
not important (see Sec. \ref{sec:weak}).
\begin{figure}[th]
\includegraphics*[width=8cm]{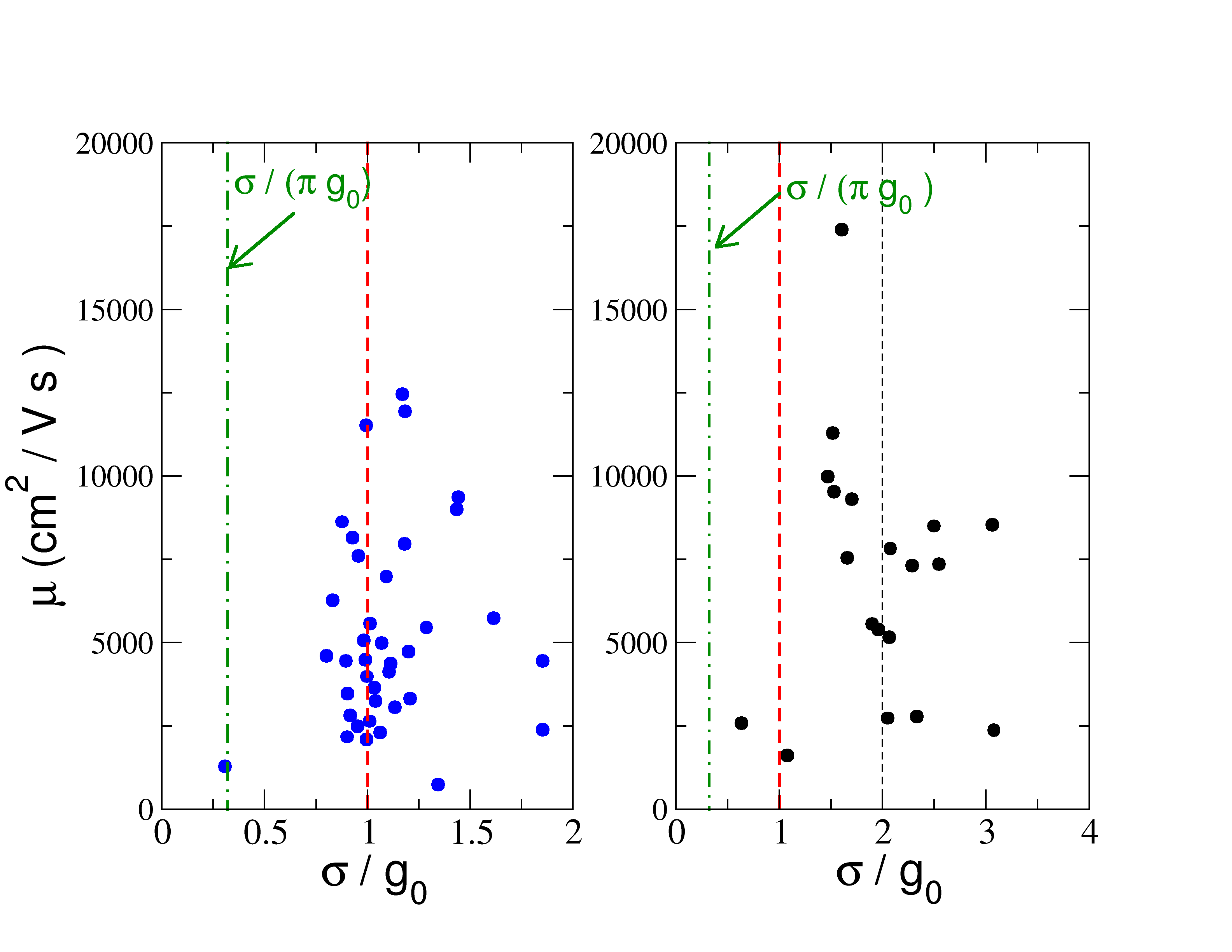}
\caption{(Color online)
The conductivity of graphene at the neutrality point.
{\bf Left:}
conductivity minimum from the Manchester's
group [Data from \cite{natMat}].
{\bf Right:} Conductivity minimum from the Columbia's
group [data from \cite{kimsigmamin}] as function of the mobility of the devices.
In both panels, several devices with different mobilities $\mu$
were measured. The mobility is defined as
$\mu=\sigma(\epsilon_F)/(e n)$, where $n$ is the electron density, and is a
measure of the
amount of disorder in the system.
The constant $g_0$ stands for twice the quantum of conductance,
$2e^2/h\simeq$0.078 (k$\Omega$)$^{-1}$, and is represented by the dashed line.
The dash-dotted line represents the value
$\sigma=\sigma_{\rm min}$, obtained in Sec. \ref{sec:disorder}.}
\label{fig:sigma_minimum}
\end{figure}

If one considers that the scattering centers are charged impurities, the 
conductivity of graphene, at the neutrality point, acquires the form
\cite{foglerPRL}
$\sigma_{\rm min}=(e^2/h)c{\cal L}$ with $c=0.5\pm0.05$ and 
${\cal L}$ the solution of the transcendent equation 
${\cal L}=\ln({\cal L}/4\alpha_g^{\rm eff})$, where $\alpha_g^{\rm eff}$ is the 
effective fine structure constant of graphene (see Sec. \ref{sec:coulomb}). 
This result for the conductivity minimum is different from that
obtained for strong short-range scatterers.

Finally,  the measured
conductivity minimum, as shown in Fig. \ref{fig:sigma_minimum}, has the same
order of magnitude as that given by $\sigma_{\rm min}$, but is larger than this
value and
has a finite variance. It is important to note that the assessment of the
transport
properties of graphene at the neutrality (or Dirac) point can be strongly
affected by the used probe geometry, the use of invasive contacts, or the
lack of effective control on the sample's homogeneity
\cite{blakemetals,smith_ele_holes}.
These effects are responsible for the differences in the two sets of
measurements
 given  in Fig. \ref{fig:sigma_minimum}.

\subsection{Calculation of the conductivity minimum for pristine
graphene ribbons}
\label{sec:ballistic}

Graphene ribbons have been produced by different methods:
 etching  of exfoliated graphene \cite{barbaros}, using chemical
reactions \cite{smoothribbonsDAI,smoothribbonsDAIi}, unzipping
carbon nanotubes \cite{ribbonsfromtubes}, and tailoring them
by scanning tunneling microscope lithography \cite{ribbonsfromSTM}.
Much of the experimental challenge regarding the production of nanoribbons
is related to the discovery of an experimental procedure allowing, in a
systematic way,
the engineering of ribbons with fixed widths and perfect edges, together with a
detailed
characterization of their transport properties \cite{kimribbons}.

The previous section addressed the problem of the conductivity minimum of
graphene
 from the point of view  of disorder.
Another relevant problem is that of the
transport properties of pristine ribbons, where  electrons can be
in the ballistic regime. The problem we are about to discuss is a rather
interesting one,
since electronic transport will proceed via evanescent modes, whereas in normal
metals
charge transport is associated with propagating states.

We show below,
and also in this case, the system has a finite conductivity, which in some
conditions
has the same value we found in Sec. \ref{sec:disorder}, although the
physical mechanism is different. The approach to the calculation
of the conductivity of ribbons in the ballistic regime
uses  Landauer's formalism \cite{quantumtransp}, where the relevant quantity to
be
computed is the conductance of the system, which can formally
take into account quantum interference effects, absent from
the elementary Boltzmann's transport theory (but see Sec. \ref{sec:weak}).

The  measurements in bulk metals of  dc-transport properties
 allows one obtain directly  the resistance $R$
of the sample, from which the linear conductance $G=1/R$ can be determined.
In bulk metals, we can define a  material intrinsic quantity, the conductivity
$\sigma$.
Taking the example of a 2D system, we have
$\sigma=G L_x/L_y$,  where $L_x$ and $L_y$
are the longitudinal and transverse dimensions of the bulk sample, respectively. The
conductivity
is a well-defined quantity whenever the system is large enough, such that the
electronic
current is homogeneous and insensitive to variations of the impurities' position
from sample to sample.  In this regime the transport is well described by
 Boltzmann's transport equation. The validity of this equation
assumes that \cite{Ferry}:
(i) the scattering process is local in space and time, (ii) the scattering is
weak and the electric field is small, and (iii) the de Broglie wavelength of the
electron at the Fermi surface
 is much smaller than the distance between impurities. The systems amenable to
such description are said to be
{\it self averaging}.
(In 2D,
both $\sigma$ and $G$ have the same units, $1/\Omega$.)

When the  system's size is reduced, we enter the realm of mesoscopic physics.
It is instructive to compute the order of magnitude of the number of impurities
in a
graphene flake with an area of $A=L^2$, and $L=0.25$ $\mu$m (see Sec.
\ref{sec:introduction}).
Taking  $N_i/A=5\times 10^{11}$ cm$^{-2}$ as a typical impurities' concentration
in graphene
(see Sec. \ref{sec:mid-gapmodel} for understanding the origin of this number),
we obtain
$N_{i}\sim 3\times 10^2$ impurities.  The typical distance between impurities
is $d\sim \sqrt{A/N_i}\sim$0.02 $\mu$m; a typical Fermi wave number for the
electrons is graphene is $k_F\sim$0.003 $\mu$m$^{-1}$ (see Sec.
\ref{sec:finiten}), from which  follows that
de Broglie wavelength of the electrons at the Fermi surface
 is $\lambda_F=2\pi/k_F\sim$0.02 $\mu$m, making $d$ and
$\lambda_F$ of the same order of magnitude.
In this regime, the current becomes non-homogeneous and sensitive to the
position of the impurities in the material. Then, the conductance
shows fluctuations from sample to sample, and the concept of conductivity
loses its meaning. Metallic systems such as graphene are considered highly
conducting
but disordered metals. The behavior of the electrons becomes sensitive
to the metal contacts, surfaces, and interfaces as well and quantum mechanical
interference effects become important.
Due to these interference effects, the transport properties of mesoscopic
systems in the ballistic regime are better assessed by the Landauer's formalism
\cite{Ferry}.

In calculating the conductance of pristine graphene ribbons,
we assume a ribbon of length $L_x$ and width $L_y$, connected
to  heavily doped (say, with electrons) graphene leads (see Fig.
\ref{fig:strain_tile} for the geometry of the device).
The
doped graphene leads will act as electron reservoirs, and the doping
is modeled by gating the leads at a potential $V_g$.

Since the leads are gated, there is a mismatch between the longitudinal
momentum $k_x$ of the electrons in the leads and in the central
part of the device, where
the undoped graphene ribbon lies; in the device
 electrons have longitudinal momentum $q_x$. The momentum $k_y$ is in this
case a conserved
quantity.
The problem is then that of computing the transmission
amplitude for an electron coming from the left lead to emerge at the right one.
The energy of the electrons at the right and left leads is given by
 $E=-eV_g\pm v_F\sqrt{k_x^2+k_y^2}$; in the central region the energy is
given by  $E=\pm v_F\sqrt{q_x^2+k_y^2}$. We further impose periodic boundary
conditions along the transverse direction,
which gives $k_y=2\pi n/L_y$ with $n=0,\pm 1,\pm 2, \ldots$.
Since we are interested in graphene's transport properties at the Dirac point,
we have to consider the case of zero energy.
For this energy, the solution
of $\sqrt{q_x^2+k_y^2}=0$ gives $q_x=ik_y$ and therefore the propagation of the
electrons in the central region proceeds by means of evanescent waves.

The scattering problem requires writing the wave function on the
left and right leads, and on the central region \cite{katsnelson,tworzydlo}.
 In the left lead, the wave function,
up to a multiplicative factor of $e^{iyk_y}$, reads
\begin{equation}
\psi_{L}(\bm r)=
\left(
\begin{array}{c}
1\\
e^{i\theta(\bm k)}
\end{array}
\right)e^{ik_xx}+
 r_n\left(
\begin{array}{c}
1\\
-e^{-i\theta(\bm k)}
\end{array}
\right)e^{-ik_xx}\,.
\label{eq:psiL}
\end{equation}
In the central region the wave function can be written as
\begin{equation}
\psi_{C}(\bm r)= a_n\left(
\begin{array}{c}
0\\
1
\end{array}
\right)e^{-k_yx}
+b_n\left(
\begin{array}{c}
1\\
0
\end{array}
\right)e^{k_yx}\,.
\end{equation}
Finally, in the right lead we have
\begin{equation}
\psi_{R}(\bm r)=
 t_n\left(
\begin{array}{c}
1\\
e^{i\theta(\bm k)}
\end{array}
\right)e^{ik_xx}\,.
\label{eq:psiR}
\end{equation}
The calculation of the transmission amounts to imposing the continuity
of the wave function at $x=0$ and $x=L_x$ and determining the
transmission amplitude $t_n$ from
which the transmission associated to a given transverse mode
$n$ is obtained as $T_n=\vert t_n\vert^2$ (to each quantized $k_y$ momentum
corresponds a $n$ transverse mode). The final result for the total transmission
at zero energy  is
\cite{katsnelson,tworzydlo}:
$
T=\sum_n T_n\simeq  \sum_n 1/\cosh^2(k_yL_x)
$.
As stated, the conductance $G$ is expressed in terms of the conductivity
$\sigma$ as $G=4e^2T/h=\sigma L_y/L_x$.
In the regime $L_y/L_x\gg 1$, corresponding to ballistic transport, we have
$T\simeq L_y/(L_x\pi)$,
and therefore $\sigma=\sigma_{\rm min}$, the same value
obtained in Eq. (\ref{eq:sigmamin}),
 due to
disorder. We  stress that, for graphene ribbons,
only in the regime  $L_y/L_x\gg 1$ is the conductivity a well defined quantity,
since only in this case is this quantity independent off the aspect ratio of the
ribbon.

The extension of this type of calculations to finite temperatures is elementary,
and it follows from the Laudauer's formalism as well.
Such theoretical investigations were done and the results seem  to be in
qualitative agreement with  transport measurements made in high-mobility
suspended graphene
\cite{SigmaTsuspended}.

The conductance of ribbons, with aspect ratio $L_y/L_x\gg 1$, was experimentally
measured,
and the value $\sigma=\sigma_{\rm min}$ was obtained \cite{miao,morpurgopi}
in agreement with the previous result.
There are, however, difficulties associated with measuring the
conductivity of graphene ribbons at the neutrality point \cite{blakemetals},
since inhomogeneous samples tend to overvalue the minimum of conductivity and
two-probe measurements are generally expected to undervalue it
\cite{blakemetals}.
Due to these subtleties, there is some reserve in the community
\cite{blakemetals}
regarding the
 measured conductances \cite{miao,morpurgopi}.

We note that the above result for $\sigma_{\rm min}$, being equal
to that computed in Sec. \ref{sec:disorder}, has  a different
physical origin. The result  obtained here is only valid in the regime
$L_y/L_x\gg 1$, when the system is in the ballistic regime.
However, one must recognize that the presence of the evanescent modes in the
above calculation produces a finite density of states at the Dirac point,
precisely what happens in the bulk  disordered graphene calculation 
discussed in Sec. \ref{sec:disorder}. When the calculation just described
for graphene in ballistic
regime ($L_y\gg L_x$)
includes the effect of resonant scatterers, the conductance is corrected by
the value $\delta G=4\sigma_{\rm min}/\pi$, per resonant scatterer
\cite{TitovRS},
that is, we have impurity-assisted tunneling \cite{Titovsingle}.

As a last comment, we  note that the important topics of
edge disorder \cite{Mucciolo,Caio,Gallagher}
and Coulomb blockade in graphene nanoribbons are not considered
in this Colloquium, since they
have been considered elsewhere \cite{Dubois}. A review 
on the effect of disorder on the electronic transport in
graphene nanoribbons is also available \cite{eduardoreview}.

\subsection{Puddles}
\label{sec:puddles}

We  now address the fact that the model developed in Sec.
\ref{sec:ballistic} for the
conductivity of graphene at the Dirac point is somewhat
simplistic, since it assumes the possibility
to have graphene with exactly zero electronic density at $E= 0$, the neutrality
point.

\begin{figure}[th]
\includegraphics*[width=7cm]{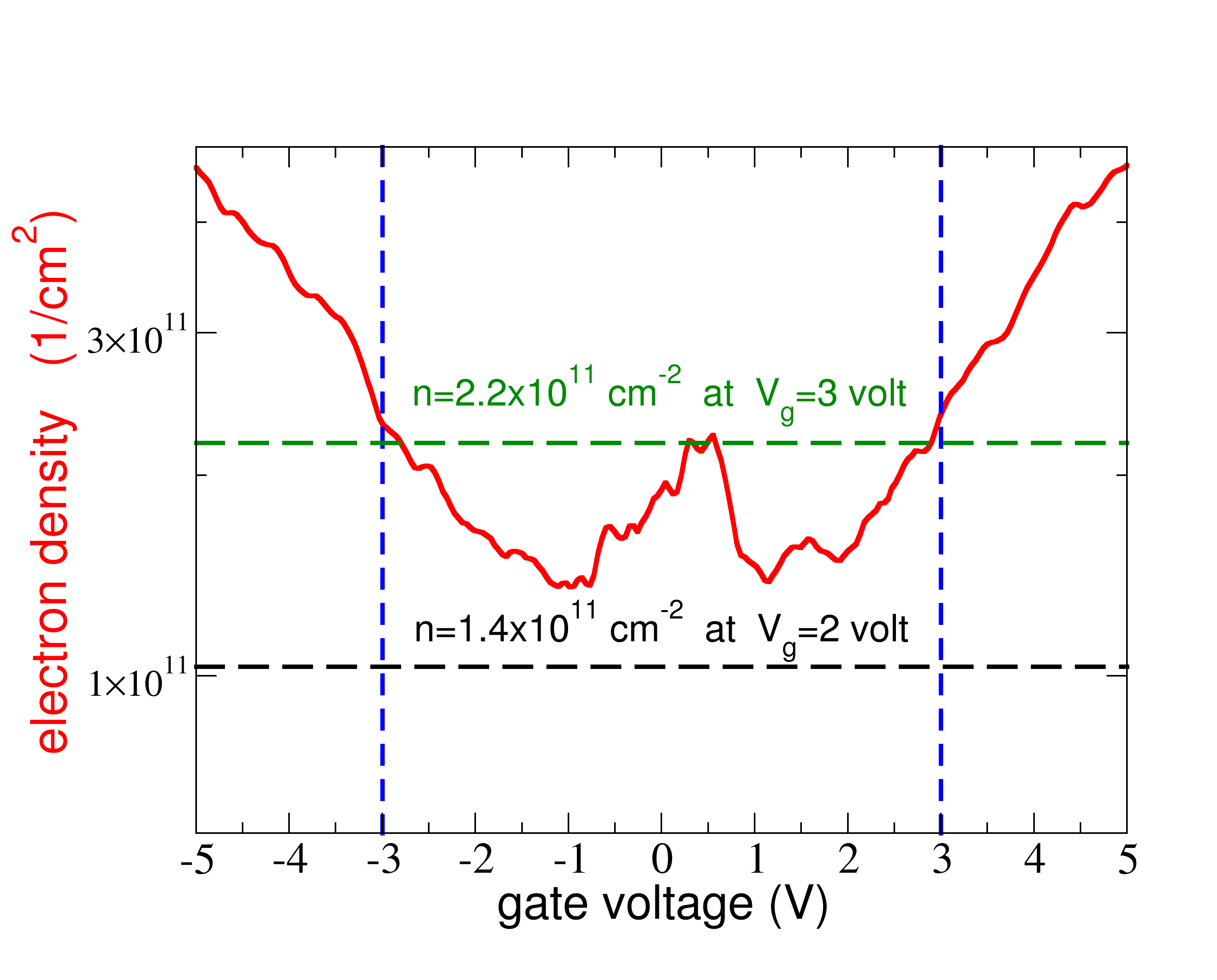}
\caption{(Color online) Dependence of the absolute value
of electron density on the gate voltage $V_g$. In Sec. \ref{sec:finiten},
we show that the electron density $n$
as function of the gate voltage  follows 
$n = 7.2\times 10^{10}V_g$ cm$^{-2}$. Using this, the electron
density for $V_g=3$ V should be $n\simeq 2.2\times10^{11}$ cm$^{-2}$ whereas
for $V_g = 2$ eV we should have $n\simeq 1.4\times10^{11}$ cm$^{-2}$. It is
clear
 that for $V_g=3$ the electronic density is the
predicted one; however
$n$ never equals its predicted value
for $V_g \lesssim 2$ V. Note that the vertical scale is logarithmic.
(Data from M. Monteverde
{\it et al.} \cite{miguelmonteverde}, courtesy of M. Monteverde.)}
\label{fig:monteverde}
\end{figure}

The physics close to the Dirac point is different
from that at finite densities (to be discussed in Sec. \ref{sec:finiten})
as  suggested by the data shown in Fig. \ref{fig:monteverde}. In Sec. \ref{sec:finiten}
we show that the electronic
density $n$  can in graphene on top of silicon oxide
be externally controlled by a gate potential
$V_g$ and given by $n\simeq  7.2\times 10^{10}\times V_g$
cm$^{-2}$.
According
to this equation, the electron density can be tuned all the
way down to zero by changing the gate potential. However, in
Fig. \ref{fig:monteverde} we show that the absolute value of the electron
density never drops below its theoretically predicted value for $V_g=2$ V.
This experimental fact hints for different physics close
to the Dirac point, where the system  shows important
charge-density fluctuations caused by the random electrostatic potential due to
 sub-surface charged impurities. We, however, stress
that in suspended annealed graphene the electronic density can be made
as low as $\sim 10^{8}$ cm$^{-2}$,  which corresponds to about a single electron
present in a
micron size device.
Additionally, graphene's topography shows corrugations,
which are
probably due to roughness
in the underlying SiO$_2$ surface and due to intrinsic ripples of the
graphene sheet.

As mentioned, the calculations  in Secs. \ref{sec:disorder} and
\ref{sec:ballistic}
assume  as the starting point that graphene is a perfectly flat material, with
null
electronic density everywhere. However, experiments using
a scanning single-electron transistor \cite{yacoby} found
that the idealized models of Secs. \ref{sec:disorder} and
\ref{sec:ballistic} do not hold.
Those investigations \cite{yacoby} found  undoped graphene to be
a non-homogeneous system, with
 electron and hole puddles coexisting, with variations in
the electronic density in the range
$n\in[-1,1]\times 10^{11}$ cm$^{-2}$,  which corresponds to
a spatial variation of the surface electrostatic-potential  in the range
$[-0.25,0.25]$ V, with a full width at half maximum of 50 mV.
The behavior shown in Fig. \ref{fig:monteverde} is an indirect signature of
this experimental fact.
The existence of
puddles renders
 the descriptions of Secs. \ref{sec:disorder} and
\ref{sec:ballistic} unsuitable.

Posteriorly, a scanning tunneling microscopy (STM) study
\cite{Cromieinhomogeneous}
was able
to provide  detailed information on the size and electronic density value
of the puddles. This study allowed one to characterize the puddles
with  electron-density spatial resolution
two orders of
magnitude higher than previous investigations \cite{yacoby}.
From the results of Sec. \ref{sec:dispersion}, we can write
a relation between the electronic density and the energy as
$n=E^2/(\pi v_F^2\hbar^2)$. Given the presence of the puddles, the energy
becomes
 function of position, as does the electronic density. We thus have a relation
between the electronic density at the Dirac point and the energy, reading
\begin{equation}
 n(x,y)= \frac{E^2_D(x,y)}{\pi v_F^2\hbar^2}\,.
\end{equation}
The STM allows the determination of $E^2_D(x,y)$, from which $n(x,y)$ is
obtained.
These studies revealed that the average lateral dimension of the puddles is of
the order of
$\sim$20 nm [a theoretical study \cite{rossi} obtained a similar value],  and
that each of these puddles contains, on average, a charge of
$0.3\pm0.2$ electron. 
A  Kohn-Sham theory of the 
carrier-density distribution of massless Dirac fermions in the 
presence of arbitrary external potentials has also predicted the 
existence of the puddles \cite{poliniA}.
It was  experimentally determined that the topographical
corrugations
in graphene
are about an order of magnitude smaller than the puddles' size, and therefore
cannot
justify their origin. Indeed, it was established that individual
sub-surface
charged impurities are responsible for the formation of the puddles. It was
estimated that the charge fluctuations associated with a single of those
impurities
is of the order of $0.07\pm0.03$ electron. There is, therefore, a consensus
that the physics of the puddles is due to charged scatterers. The origin of such
charged scatterers is likely to be due to chemical species physisorbed onto
graphene,
which have been trapped in between
the substrate and the graphene sheet during fabrication process of the device.

From a theoretical point of view,
graphene in the puddles regime can be thought as a random resistor network
\cite{cheianovrandom}. Since Klein tunneling \cite{beenakkrmp}
is exponentially suppressed if
the barriers are not perfect potential steps \cite{cheianovKlein},
a large electronic transmission
will not occur, except for perfectly normal incidence
[things are markedly different
for magnetic barriers as opposed to electrostatic ones \cite{KleinMagnetic}];
the essential physics relating the smoothness of potential barriers to the   
suppression of  Klein tunneling was studied in the early days of relativistic quantum 
mechanics, following a suggestion by Bohr \cite{Sauter,Paolo}.
The validity of the random resistor model depends on the assumption that
transport is  incoherent at scales larger than the puddle sizes.
Due to Klein tunneling, massless Dirac electrons cannot also
localize (Anderson localization)
\cite{klein,cheianovKlein,Mucciolo,Caio} under the effect of the random
electrostatic
potential (long-range scatterers) creating the puddles; this accounts for the
finite conductivity of
graphene at the Dirac point.
As discussed  in Sec. \ref{sec:weakgraph}, long range scatterers
preclude the possibility of weak localization effects, and since the
electrostatic potential variations
can be attributed to charged impurities, the description of transport at the
Dirac point based
on such type of scatterers seems to be the correct approach \cite{sarna2pnas}.

We  note that intra-cone
backscattering
(see Sec. \ref{sec:weakgraph})
has been shown to be present in graphene   \cite{Cromieinhomogeneous},
which in view of Klein tunneling is a rather interesting experimental fact.
Finally, when strong inter-valley scattering is present, electrons in graphene
can
localize.

\section{The transport properties of graphene at finite electronic density}
\label{sec:finiten}

In the previous section we  discussed STM experiments
 supporting the theory \cite{rossi,poliniA} that charge scatterers dominate the
electronic transport of neutral graphene. In the ensuing sections, we 
discuss
transport in doped graphene, analyzing the role that resonant and charged
scatterers
play in this regime.

\subsection{The dependence of the conductivity on the
gate voltage}
\label{sec:gate}

We now discuss the dependence of the conductivity of
graphene on the gate voltage,  considering two
different types of scatterers:
resonant scatterers (strong short-range scatterers)
and charged impurities.
We shall not discuss here scattering from random strain \cite{kastripples},
which we defer to Sec. \ref{sec:strain}.

The electronic density in graphene can be controlled by the back-gate of
a device  engineered as a plane capacitor -- a field effect transistor, made of
silicon
oxide (relative
permittivity $\epsilon=3.9$), with a thickness
$b$ of $\sim$300 nm.
According to elementary electrostatics, the electric field
in the dielectric is given by $E_{\rm cap}=en/(\epsilon_0\epsilon)$, with
$n$ the surface electronic density of graphene, which acts as one plate of the
capacitor.
The gate potential is related to the electric field by $E_{\rm cap}=V_g/b$,
and so
the density of induced charge is $n=\epsilon_0\epsilon V_g/( e b)$.
Inserting the numerical values of $\epsilon$ and $b$, we obtain $n=\alpha V_g$,
with
$\alpha\simeq 7.2\times 10^{10}$ V$^{-1}\cdot$cm$^{-2}$.
The Fermi momentum $k_F$ is obtained via
$k_F=\sqrt{\alpha\pi V_g}$, a result derived by
counting the states in momentum space up to $k_F$.

Ever since the original paper on graphene \cite{nov04}, demonstrating
the ambipolar field effect, it became clear that the conductivity of graphene
depends on the gate voltage in some circumstances   roughly
 as
$\sigma(\epsilon_F)\propto V_g$; this is shown in  Fig.
\ref{fig:chargescatterers},
after some replotting of the data  (solid curve on the right panel).
Experiments also show conductivities presenting a sub-linear behavior;
see   Figs. \ref{fig:suspended} (solid curves) and
Fig. \ref{fig:chargescatterers} (dashed curve on the right panel).
Mobilities, a measure of the quality of the electronic transport
(see caption of Fig. \ref{fig:sigma_minimum} for the definition of the mobility $\mu$),
as high as $\mu\sim1\times10^7$ cm$^2\cdot$V$^{-1}\cdot$s$^{-1}$,
have been indirectly measured by Landau level spectroscopy
\cite{LiandreiPLR,Liandrei}
of graphene flakes
on top of graphite \cite{howperfect}, raising the question of how perfect can
graphene
be \cite{howperfect}. Ultimately, the answer requires the identification
of the limiting sources of electronic scattering in graphene
(among those listed in Sec. \ref{sec:sourcesdisorder}).

An approach combining Fermi's golden rule, Boltzmann equation,
the Coulomb potential created by screened charged impurities, and
a random phase approximation calculation of the dielectric function of graphene
\cite{shung,Wunsch:2006} gave a
first good account of the observed $\sigma(\epsilon_F)\propto V_g$
behavior  for graphene's conductivity \cite{sarna1,sarna2pnas,sarna3}.
When graphene was doped with potassium \cite{fuhrer},
the measured conductivity agreed with the theory \cite{sarna2pnas}, as expected.
[We  note that the conductivity of graphene covered by
metal clusters it is still far from being fully understood \cite{piclusters}.]

Using the same approach for a delta-function potential
\cite{sarna1,sarna2pnas,sarna3},
the prototype of a short range scatterer,
the computed conductivity is a constant number, independent of the gate voltage
and of the dielectric constant of the medium. 
In what follows, we argue that this result is inconsistent.
We note  that an attempt
to solve the Lippmann-Schwinger equation for a delta-function potential showed
that this
problem is ill defined (regularization of the problem is required in order to have
a well-defined problem; as usual, this procedure introduces a length scale.
 This length scale is interpreted as the range of the short-range potential), 
and therefore the first Born approximation cannot be
trusted. Indeed, exact numerical calculations show that the 
first Born approximation is inadequate to describe the role of strong short-range
scatterers in graphene \cite{Igor}.  
At the same time,
a numerical calculation using the Kubo-Greenwood formalism \cite{nomura}
showed that  $\sigma(\epsilon_F)\propto V_g$ for
charged impurities (the level broadening due to scattering
was however introduced by hand).
The same work \cite{nomura} also showed
that short-range impurities do produce
a conductivity that depends on the gate voltage, but in a sub-linear manner.
Also, previous calculations of $\sigma(\epsilon_F)$ based on the SCBA showed
that strong short-range scatterers, described by delta-function potentials, do
give rise
to a gate-voltage-dependent conductivity
\cite{nmrPRB06,andoshon}, a result embodied in Eq.
(\ref{eq:sigmamin}). A similar conclusion was obtained from
a semi-classical approach taking into account the
chiral nature of massless Dirac fermions \cite{schliemann}.

The two different  results -- those based on
Fermi's golden rule, as opposed
to those obtained from the SCBA, for strong short-range potentials -- are easily
understood:
the SCBA is a non-perturbative method, suitable for strong short-range
potentials, which takes into account the large deviation of the wave function,
within the potential range, from the usual plane wave  used
in the first Born approximation, as pointed out by Peierls  \cite{peierls}: 
indeed, the first Born approximation produces a large
scattering cross
section, whereas the exact calculation gives a small value. Since the
conductivity depends on the
scattering (transport) cross section, an incorrect determination of it will
give, at least,
an incorrect value for the impurity concentration in the material.
Unfortunately, the reliance on the result based on Fermi's golden
rule is widespread in the community and is being used to
fit the experimental data \cite{sarnafit}, at the same time that the
resonant scattering mechanism points toward the presence in the material
of strong localized potentials (see Sec.\ref{sec:sourcesdisorder}).

If it is certain that some amount of charged impurities is
present at the silicon oxide--graphene interface (responsible for the
electron and holes puddles), it is no less true that
 experiments do not rule out other
 sources of scattering. Indeed, recent experiments showed that
both adsorbed hydrogen and vacancies led to  conductivity curves
indistinguishable in form from those of pristine graphene \cite{Chen,Elias}.
The presence of these short-range scatterers -- vacancies and hydrogen --
is signaled by a
significant Raman $D-$band intensity \cite{Chen}, since they
 couple electron states from
the $\bm K$ and $\bm K'$ valleys (see Sec. \ref{sec:weakgraph}).
By the same token,
the presence in pristine  graphene
of such $D-$band would be the signature of the presence of short-range
scatterers
in the material. Detailed Raman investigations in pristine graphene
have  been carried out \cite{Dpeak}, showing that, indeed,
 a small $D-$peak is present in the Raman spectrum of the pristine material.
Subsequent transport experiments \cite{Dpeak}
support strong
 short-range scatterers as the limiting
source of scattering in graphene.

An experiment especially designed to address the importance of
charged impurities used devices with dielectrics having high permittivity
constants \cite{geimhighK}. These experiments
did not  exclude completely the contribution of
this type of impurities, but did challenge the idea that
charged impurities are the main source of scattering in graphene.

On the other hand, in another set of experiments, an apparently similar
investigation was done,
but with ice layers on top of graphene  and reaching a different conclusion.
It was argued that the results were consistent with charge scattering
\cite{Fuhrerice}.
There is, however, at least one difficulty with the arguments developed
in that work: the number of ice atomic layers was at the most six and therefore
can hardly be considered an infinite dielectric made of ice; the lines of the
electric
field are essentially in the vacuum \cite{slab1,slab2}.

A number of questions can still be asked
\cite{geimhighK,shedin,miguelmonteverde}:
\begin{enumerate}
 
\item
In a study of graphene's sensitivity to gases \cite{shedin} (NO$_2$,
H$_2$O, and iodine acting as acceptors, whereas NH$_3$, CO,
and ethanol acting as donors), chemically-induced charge-carriers
 concentrations as large as $50\times 10^{10}$ cm$^{-2}$ were achieved.
The induced
chemical doping  shifted only the neutrality point of the conductivity curves,
without any significant changes either in  the shape of those curves
or in the mobility of the devices; the estimated
concentrations of added charged scattering centers was high as $10^{12}$
cm$^{-2}$
 \cite{shedin}.
Why is it that no appreciable changes in the mobility were measured
in these experiments? [One possible way out can be envisioned:
the chemical dopants may cluster, and this would reduce the effectiveness
of their scattering effect \cite{kastclusters}.] We  also note
that the definition of the mobility used in the analysis of the
data \cite{shedin} has been criticized
in the literature \cite{adsorbedmolecules}.

\item In a study designed to test the prediction \cite{sarna3}
of the
charge scattering model
for the ratio  of the
transport scattering time $\tau$ and the elastic scattering
time $\tau_e$  for both
monolayer and bilayer graphene,
the experiments found  disagreement between the predicted behavior and the
measured data, for both graphene systems.  The measured deviations
were found to be stronger for bilayer graphene \cite{miguelmonteverde}. Further,
it was found
that the measured data agree with the resonant scattering mechanism.
How to reconcile this set of measurements with models explaining the mobility
of both monolayer and bilayer graphene based on the charge scattering mechanism
\cite{avourisscattering}? [We mention that Monteverde's
{\it et al.}
results must be confronted with those of a similar experiment \cite{sarnafit},
reaching different conclusions.]

\item  Since screening is strongly dependent on the value of the permitivity
$\epsilon$
of the surrounding medium,
why is the mobility almost insensitive to changes of this parameter? For
example,
$\epsilon$ for ethanol changes from 25 to 55 as the temperature
drops  from 300 K down to $\sim$160 K, but an experiment done in ethanol showed
no variation of graphene's mobility. We, however,
note that in some experiments \cite{geimhighK}
a certain amount of variation in the mobility was measured in some devices
upon changing  the dielectric
constant. This result does show that charged impurities play some role
as scattering centers but apparently not the
limiting  one.  
  
\end{enumerate}

The answers to the above questions remain debatable to some extent.
The clarification of some of these issues could be taken to an ultimate
test  using a solid dielectric with a  high relative permittivity.
It just happens that strontium titanate (SrTiO$_3$) has a
relative permittivity of about 10,000 below $T=50$ K, which
suddenly drops to 300  when the temperature rises above 50 K. A device
 using such a dielectric would produce a dramatic change
of the mobility upon a  drop in temperature  below  50 K.

If we now refocus our attention on the role of strong
short range scatterers,
we recall that both the linear and sub-linear
behaviors can be accommodated within a model
based on what is now called {\it resonant scatterers dominated
conductivity}, giving
rise
to  mid-gap states
\cite{robinson,stauberBZ,basko,wehling,wehlingII}, plus the additional
effect of charged impurities, which, however,
do not play the central role.
On the other hand, the simplest model based on short range scatterers,
in which the effect
of charged impurities is ignored, does not account for the observed
dependence of the mobility on the dielectric constant of the device, which has
been
shown experimentally to be present to some extent \cite{geimhighK}.
In Sec. \ref{sec:mid-gapmodel}
we present the main results of such a simple  model, and in Sec.
\ref{sec:coulomb}
we include the role of charged impurities, and an improved model
taking into account both types of scatterers is given. This latter model is a
simple combination of results already available in the literature, albeit
presented with a different emphasis.

We   stress that
from the analysis of the SCBA results  we see
that features
showing  at energies close to the Dirac point
are all proportional to $\sqrt{n_i}$ (recall that $n_i$ is the
density of impurities per unit cell), as shown in Fig. \ref{fig:CPA}.
This introduces an energy scale
$\epsilon_{\rm min}\lesssim \hbar v_F \sqrt{n_i}/a_0$,
below which electron scattering based on plane waves breaks down, meaning that
close to the Dirac point the results of Secs. \ref{sec:mid-gapmodel} and
\ref{sec:coulomb} are expected not to hold.

 Both models based on mid-gap
states or on charged impurities, presented below, fail to give a
satisfactory account
of the physics close to the Dirac point, since they are based on the scattering
of plane waves.

Finally,
in suspended graphene \cite{evasuspended,kimsuspended,bolotinsuspended},
where the material is hanging over
a trench,
mid-gap states are expected to survive, since some fraction of the
corresponding scatterers will still be present,
whereas charged impurities are expected to be absent.
In the suspended situation,
the mobility
of the material will be limited only by
resonant scatterers
plus ripples induced by strain due to the  electric field created by the
gate \cite{bao,Fogler:2008}.

\subsection{Partial-wave description of resonant scatterers}
\label{sec:mid-gapmodel}

We assume the presence of
short-range scatterers, which we model here as
 disks of radius $R$, and whose origin was discussed in
Sec. \ref{sec:disorder}. The effect of the scatterers
is  such that the electron wave function
is zero for $r<R$. The sizes of these disks are of the order
of the size of the primitive lattice vectors.
The circular shape takes the isotropy of the scattering process into a account,
and the boundary condition allows a simple analytical solution.
A vacancy is one of the possible physical realizations of the model  just
introduced.

The Dirac Hamiltonian (\ref{eq:dirac}) in polar coordinates $r$ and $\varphi$
reads
\cite{hentschel,rodrigues,recher}
\begin{equation}
H_{\bm K}=-iv_F\hbar\left(
\begin{array}{cc}
0 & L_-\\
L_+ & 0
\end{array}
\right) \,,
\label{eq:Diracpolar}
\end{equation}
with $L_\pm=e^{\pm i\varphi}(\partial/\partial r \pm ir^{-1}
\partial/\partial\varphi)$.
A particular solution of Eq.
(\ref{eq:Diracpolar}) with eigenvalue $v_F\hbar k$
has the form $\psi_m(rk)\propto (J_m(kr)e^{-im\varphi},
-iJ_{m+1}(kr)e^{-i(m+1)\varphi})^\dag$, where $J_m(z)$ is the regular Bessel
function
of first kind
and integer order $m$. This solution corresponds to a partial
wave in the angular momentum representation of the plane wave. In the presence
of
the potential created by the disk we write the trial wave function as
\begin{eqnarray}
\psi(rk) = A
\left(
\begin{array}{c}
J_m(kr)e^{im\varphi} \\
iJ_{m+1}(kr)e^{i(m+1)\varphi}
\end{array}
\right)\nonumber\\
+B\left(
\begin{array}{c}
Y_m(kr)e^{im\varphi} \\
iY_{m+1}(kr)e^{i(m+1)\varphi}
\end{array}
\right)\,,
\end{eqnarray}
where
$Y_m(z)$ is the irregular Bessel function
of first kind
and integer order $m$.
We consider that at $r=R$ the wave function satisfies the zig-zag boundary
conditions \cite{Adame,akhmerov}, $\psi_{1,m}(kR)/\psi_{2,m}(kR)=0$, where
$\psi_{i,m}(kr)$, with $i=1,2$, is the $i-$component of the Dirac spinor.
This boundary condition makes sense since, as discussed in Sec.
\ref{sec:sourcesdisorder},
a resonant scatterer can  effectively behave as a vacancy; in turn, a vacancy
is a three-site zig-zag edge. This boundary condition also represents
the limiting case where the electronic probability flux is zero through the
region where the potential is finite.
The phase shift $\delta_m(kR)$ of the $m$ partial wave is given by
 \cite{stauberBZ,basko,wehling,hentschel,katsnov}
\begin{equation}
\tan\delta_m(kR) =\frac{J_m(kR)}{Y_m(kR)}\,.
\end{equation}
The relative importance of the several phase-shifts to the transport scattering
cross section $\sigma_T(kR)$ depends on the value of $k_FR$. In 2D, the
differential cross section, $\sigma(\varphi)$, reads $\sigma(\varphi)=\vert
f(\varphi)\vert^2$,
with $f(\varphi)$ given by
\begin{equation}
f(\varphi)= \sqrt{\frac{2i}{\pi k }}
\sum_{m=-\infty}^{\infty}e^{i\varphi m}e^{i\delta_m(k)}\sin[\delta_m(k)]\,.
\end{equation}
When $k_FR<1$,
the $s-$wave phase shift $\delta_0(k_FR)$ is the dominant contribution.
Making use of the relation $1/\tau( k)=n_iv_F\sigma_T(kR)/A_c$ \cite{ziman},
where $\tau(k)$ is the transport relaxation time (see also Sec. \ref{sec:weak}),
and since the total transport cross section $\sigma_T(kR)$ is obtained from
\begin{equation}
\sigma_T(kR)=\int_0^{2\pi}d\,\varphi  \sigma(\varphi) (1-\cos\varphi)\,,
\end{equation}
the conductivity $\sigma(k_F)$
is given by (both spin and valley degeneracies included)
\begin{equation}
\sigma(k_F)=e^2v_F^2\frac{\tau(k)}{A_c}\rho(\epsilon_F)=
e^2v_F\frac{\rho(\epsilon_F)}{n_i\sigma_T(k_FR)}\,,
\end{equation}
with $\sigma_T(k_FR)=4\sin^2\delta_0(k_FR)/k_F$. Since we  assume
$k_FR<1$, we also have $1/\delta_0(k_FR)\simeq2\ln(k_FR)/\pi$, and the final
result for the dc-conductivity is then
\begin{equation}
\sigma(k_F)=g_0\frac{3\sqrt{3}}{4\pi}\frac{a_0^2\alpha V_g}{n_i}\ln^2(\sqrt{\alpha\pi
V_g}R)\,.
\label{eq:sigma-midgap}
\end{equation}
The result (\ref{eq:sigma-midgap}) for the conductivity holds as long
as the Fermi momentum is larger than $k_F\gtrsim\epsilon_{\rm min}/(\hbar v_F)$
(recall previous discussion).
The conclusion is that resonant scatterers (strong short-range
impurities),
giving rise to mid-gap states,
 give a conductivity that is
gate voltage dependent, with sub-linear or quasi-linear dependence
on $V_g$, depending on the size of the scattering disk $R$.
Furthermore, the conductivity (\ref{eq:sigma-midgap}) is not
independent of $V_g$ even if we take $R$ to be of the order
of the carbon-carbon distance, $a_0$.

In the first experimental study of the conductivity of suspended
graphene \cite{evasuspended}, it was shown that this quantity  
is well described by fitting it to a model of mid-gap 
scattering states, that is Eq. (\ref{eq:sigma-midgap}).

\begin{figure}[th]
\includegraphics*[width=9cm]{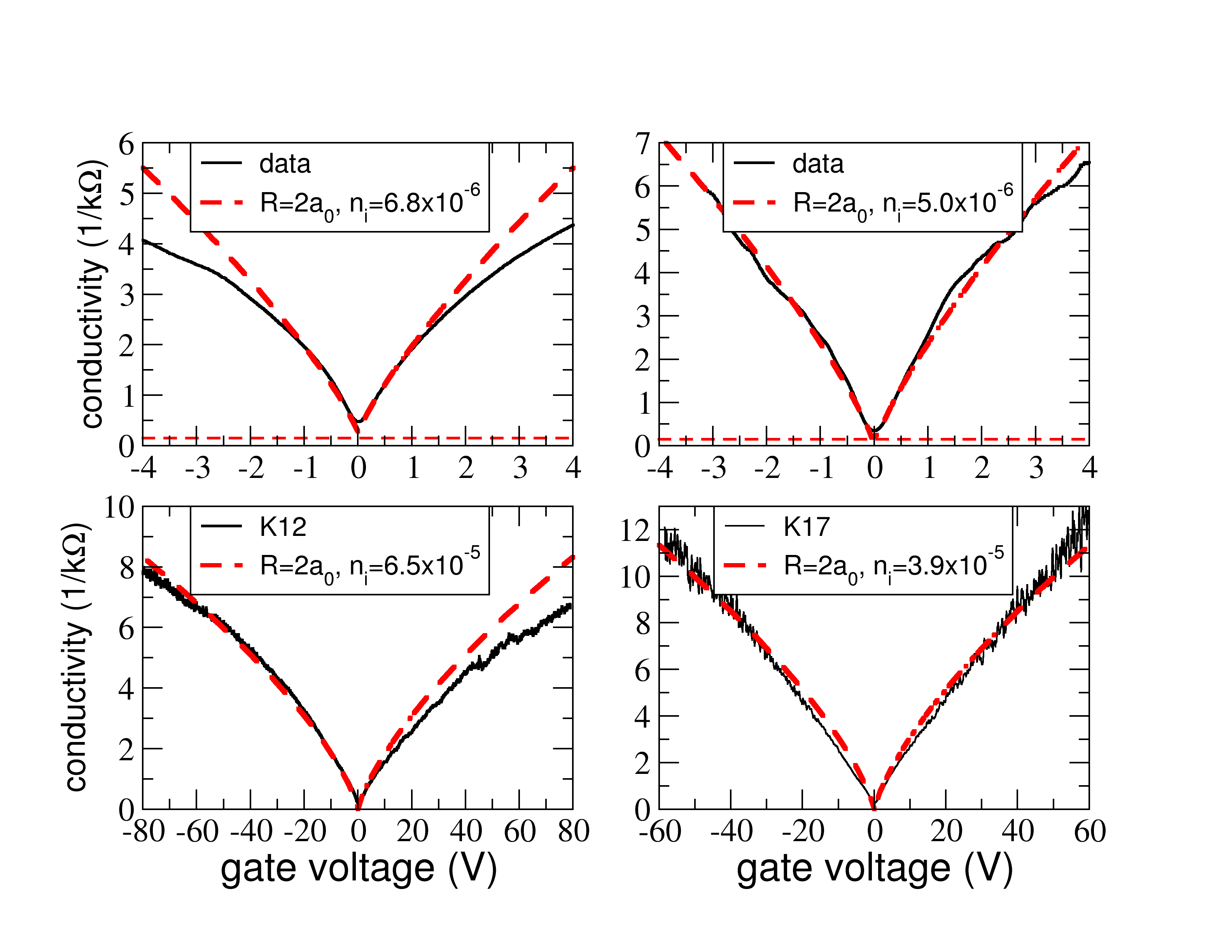}
\caption{(Color online)
The conductivity of suspended and non-suspended graphene
as function of the gate voltage.
{\bf Top panels:} Conductivity of suspended graphene from
two different devices, after current annealing ($\ell\sim$1 $\mu$m). The {\bf
top left} panel
corresponds to a device with
$\mu\sim170,000$ cm$^2\cdot$V$^{-1}\cdot$s$^{-1}$, at electronic density
of $n=2\times 10^{11}$ cm$^{-2}$ [data from K. I. Bolotin {\it et al.}
\cite{kimsuspended}, courtesy of K. I. Bolotin]. The {\bf top right} panel
corresponds to a device with
$\mu\sim200,000$ cm$^2\cdot$V$^{-1}\cdot$s$^{-1}$, at electronic density
of $n=2\times 10^{11}$ cm$^{-2}$ [data from K. I. Bolotin {\it et al.}
\cite{bolotinsuspended}, courtesy of K. I. Bolotin].
{\bf Bottom panels:} Conductivity of graphene on top of silicon oxide,
corresponding to devices with $\mu\sim10,000$ cm$^2\cdot$V$^{-1}\cdot$s$^{-1}$
(Data from  Y.-W. Tan {\it et al.}
\cite{kimsigmamin}, courtesy of P. Kim). In all panels the fits
use the model developed in Sec. (\ref{sec:mid-gapmodel}), with $n_i$ as the only
fitting parameter;
$R$ has to be of the order of $a_0$ \cite{wehlingII}. In the legends, the
concentration of impurities
is per unit cell.
}
\label{fig:suspended}
\end{figure}

In Fig. \ref{fig:suspended}, we fit  the conductivity
data of suspended and non-suspended graphene using Eq. (\ref{eq:sigma-midgap}).
In all cases a good fit is obtained, using impurity densities
ranging from $1.3\times 10^{11}$ cm$^{-2}$ for non-suspended samples
down to $0.7\times 10^{10}$ cm$^{-2}$ for suspended ones. We note that for the
devices termed $K17$ and $K12$ in Fig. \ref{fig:suspended} the
 $n_i$ values used are $n_i\simeq 0.8\times 10^{11}$ cm$^{-2}$ and
$n_i\simeq 1.3\times 10^{11}$ cm$^{-2}$, respectively.
The attempt to fit the same data with charged scatterers
\cite{sarna1,sarna2pnas,sarna3} gave
concentrations of $2.2\times 10^{11}$ cm$^{-2}$ and
$4.0\times 10^{11}$ cm$^{-2}$ for $K17$ and $K12$, respectively.
The two set of numbers for $n_i$ have the same order of magnitude,
but Eq.  (\ref{eq:sigma-midgap}) gives a better fit to the data, except
for the measurements shown in the top left panel of Fig. \ref{fig:suspended}.
For this particular device, made of suspended graphene, the
maximum of the measured conductivity is about two times smaller than that
measured in the other suspended device, whose data are shown in the
top right panel of Fig. \ref{fig:suspended},  suggesting that
additional sources of disorder may have been
introduced during the fabrication process.
We  stress that the fits done in Fig. \ref{fig:suspended},
use the density of impurities as the only fitting parameter since $R$ has to be
of the
order of $a_0$ \cite{wehlingII};  changes of order one in the value of $R$ give
 small changes for the impurity concentration. 

Some phenomenological approaches \cite{morozov}
have tried to reconcile the measured
sub-linear behavior of the conductivity of graphene with the linear behavior
(upon $V_g$)
predicted by the charge scatterers model.
In the case of the data given in the right panel of Fig. \ref{fig:suspended},
a sub-linear behavior of the conductivity upon $V_g$ (for graphene on top of
SiO$_2$)
is evident. A linear behavior could be recovered by
defining the measured resistivity $\rho_{\rm measured}$ as a sum of two terms
$\rho_{\rm measured}\equiv\rho_{g}+\rho_S$, where $\rho_S$ is a fitting
parameter.
The conductivity $\sigma_{\rm sub}=1/\rho_{\rm measured}$ is sub-linear in
$V_g$,
whereas the conductivity $\sigma_{\rm lin}=1/(\rho_{\rm measured}-\rho_S)$ shows
linear behavior.
The fitting parameter $\rho_S$ was assumed to be independent of $V_g$, and was
attributed to
short-range scatterers. The discussion presented above for  strong short range
scatterers
 showed that
this type of disorder does produce sub-linear behavior of the conductivity upon
$V_g$,
even in the case $R\sim a_0$, with the same order of magnitude for
impurity concentration as those proposed
by the charge scattering mechanism \cite{sarna1,sarna2pnas,sarna3}. It is 
obvious then that,
using the curves calculated with the mid-gap state mechanism,
it is still possible to obtain a linear dependence of the conductivity on $V_g$
by assuming
a $\rho_S$  fitting parameter as  done in the phenomenological approach
\cite{morozov}.
From the discussion  in this section it is fair to say that the
origin of $\rho_S$
still needs clarification. Moreover, we can even ask the question whether the parameter
$\rho_S$
is really needed for the interpretation of the data.

Finally, we note that in our calculation we have not included 
the effect of inter-valley scattering, which is known to be present 
when the scatterers are short-range. If that effect is included,
a contribution to the conductivity of the form given by Eq. (\ref{eq:sigma-midgap}) is found,
albeit with a different numerical prefactor \cite{OstrovskySCBA}.

\subsection{Partial-wave description of Coulomb scatterers}
\label{sec:coulomb}

We  now derive the contribution
to the conductivity of graphene due to Coulomb scatterers.
We can think of three alternative scenarios for the origin of
Coulomb scatterers: either they  exist independently of the
resonant scatterers, or the latter can themselves be charged,
carrying a  fraction of the unit charge, or both cases can coexist. Experiments
aiming
at studying in detail the Raman $D$ peak of pristine graphene, can also
shed light on this aspect, by studying suspended pristine
graphene before and after annealing.

 The solution of the Dirac equation in 2D for the
Coulomb potential  was obtained more than ten years
ago \cite{lin} for the sub-critical regime (see below),
and rediscovered in the context of graphene by
different groups \cite{coulombpereira,coulombnovikov,coulomblevitov,novikovapl},
who also solved the case of the super-critical regime.

It was known for some time \cite{linSchrod} that the
Coulomb problem in the 2D Schr\"odinger equation provides
a total cross section which does not coincide with
 the first Born approximation.
The same happens with the 2D Dirac equation
\cite{lin,coulombpereira,coulombnovikov,coulomblevitov}.
The discussion given above, restrains us from accepting
results based on the first Born approximation without a critical
analysis.

For the Coulomb problem, the Hamiltonian has the form $H=H_{\bm K}+\bm I
Ze^2/(4\pi\epsilon_0r)$
($\bm I$ a $2\times 2$ identity matrix)
and the
solution is sought in the form
\begin{equation}
\psi_j(\bm r)=
\frac{1}{\sqrt r}
\left(
\begin{array}{c}
f_j(rk)e^{i\varphi(j-1/2)}\\
\pm i g_j(rk)e^{i\varphi(j+1/2)}
\end{array}
\right) \,,
\label{eq:trialwavecoulomb}
\end{equation}
with $j$ a half-integer number.
When the trial wave function (\ref{eq:trialwavecoulomb}) is inserted into the
Dirac equation we get
\begin{equation}
\left[
\begin{array}{cc}
\epsilon +g/r & -\partial_r - j/r\\
\partial_r - j/r  & \epsilon +g/r\\
\end{array}
\right]
\left[
\begin{array}{c}
f_j\\
\pm i g_j
\end{array}
\right]
=0\,,
\label{eq:firstordercoulomb}
\end{equation}
where $j=m-1/2$, $\epsilon=E/(v_F\hbar)$,  $E$ is the energy,  $g=Z\alpha_g$,
and $\alpha_g=e^2/(4\pi\epsilon_0v_F\hbar)\simeq 2.2$ is graphene's fine
structure constant.
The solution to Eq. (\ref{eq:firstordercoulomb}) has been obtained by
several, and the central quantity
is the phase shift of the $j$ partial wave,
as in the case of Sec. \ref{sec:mid-gapmodel}.
For this problem the phase shifts
read
\begin{equation}
e^{2i\delta_j(g)}=\frac{j\Gamma(s-ig)}{\Gamma(s+1+ig)}e^{i\pi(j-s)}\,,
\label{eq:coulombphaseshift}
\end{equation}
with the property $\delta_j(g)=\delta_{-j}(g)$,
$s=\sqrt{j^2-g^2}$, and $\Gamma(x)$ the usual gamma function; the sub-critical
regime is defined
by the condition $g<1/2$. Contrary to the
short-range scatterer problem, solved in Sec. \ref{sec:mid-gapmodel},
the phase shifts (\ref{eq:coulombphaseshift})
do not depend on the energy of the incoming particle, but they do
depend on the sign of the Coulomb potential and on which type of particle,
an electron or a hole, is being scattered. The independence of $\delta_j(g)$
on the energy is simple
to understand from a straightforward argument based on dimensional
analysis: on the one hand, the Coulomb potential has no intrinsic length scale,
and on the other hand, the particle's mass is null. These two facts
show that the problem as a whole has no intrinsic length scale (some sort of {\it Bohr's
radius}, as in the non-relativistic theory of the hydrogen atom)
and therefore
dimensionless numbers involving the momentum $k$ cannot be formed, leading to
the conclusion that
the dimension
of the cross section (dimension of length, in 2D) can only come
from the momentum itself. That is,
we are then bound to have $\sigma_T(k)\propto 1/k$,
which gives the linear dependence on the gate voltage.
The electron-hole asymmetry of the cross section
can partially account for the measured asymmetry of the conductivity curves.
Another source of electron-hole asymmetry of the conductivity
is  originated in the metal contacts of the transistor
\cite{Goldhaber}.

The above solution assumes that graphene is floating in vacuum.
In a real experiment, graphene is on top of a dielectric,
SiO$_2$ being the most common. Other dielectrics have also been
used \cite{geimhighK}. In these experimental conditions, the
value of $g$ is different from that given above.

In the case $Z=1$ and for graphene
on top of a dielectric,  a charge $e$ in between the dielectric (with
relative permittivity $\epsilon_{\rm d}$) and graphene, behaves effectively as
a charge with a value \cite{Landau-Electrodynamics,slater-Electrodynamics}
 of $e_{\rm r}=2e/(1+\epsilon_{\rm d})$.
Additionally, the relative pertimittivity of graphene due to electron-electron
interactions is renormalized to $\epsilon_{\rm
r}=1+2\pi\alpha_g/[2(1+\epsilon_{\rm d})]$
\cite{sarna2pnas,shung,gonzalez}.
These two effects combined give an
effective fine structure constant for graphene of
\begin{equation}
\alpha_g^{\rm eff}= \frac{\alpha_g}{\epsilon_{\rm r}}\frac{2}{1+\epsilon_{\rm
d}}\,.
\end{equation}
Using the same procedure of Sec. \ref{sec:mid-gapmodel}, the conductivity
of graphene due to Coulomb scatterers reads
\begin{equation}
\sigma=e^2v_F\frac{k_F\rho(\epsilon_F)}{2n_i\Lambda(g)}=
g_0\frac{3\sqrt{3}a_0^2}{8n_i\Lambda(g)}\pi\alpha
V_g\,,
\label{eq:sigmaCoulomb}
\end{equation}
where
$\Lambda(g)=\sum_{m=-\infty}^{\infty}\sin^2(\delta_{m+1/2}-\delta_{m-1/2})$.
It is worth stressing that $\Lambda(g)$ is different for particles, $g>0$,
and holes, $g<0$, a behavior not captured by the first Born approximation
\cite{sarna2pnas}.
In Table \ref{tab:sigmacoulomb}
we give the numerical values for the quantity $\Lambda(g)$,
considering graphene on top of
or submerged in   different
dielectrics.

\begin{table}
\begin{tabular}{c|cccc}
\hline\\
Dielectric&$\epsilon_{\rm d}$&$\alpha_g^{\rm effct.}$& $\Lambda(g)$&
$\Lambda(-g)$\\
\hline\\
H$_2$O         &80& 0.05 & 0.013 & 0.012\\
Ethanol (160 K)&55& 0.07 & 0.027 & 0.022\\
Ethanol (300 K)&25& 0.13 & 0.10  & 0.07\\
HfO$_2$        &25& 0.13 & 0.10  & 0.07\\
SiO$_2$        &4 & 0.37 & 1.20  & 0.46\\
\hline
\end{tabular}
\caption{Dependence of $\Lambda(g)$ on the type of dielectric
for both electrons, $g>0$, and holes, $g<0$. The impurities are assumed
to have valence $Z=-e$. For ethanol, the dielectric function
depends on temperature, as indicated between braces. Since
graphene is in the ultra-relativistic limit, the value for $\Lambda(g)$
cannot be obtained from adding only few partial waves.}
\label{tab:sigmacoulomb}
\end{table}

\begin{figure}[ht]
\includegraphics*[width=9cm]{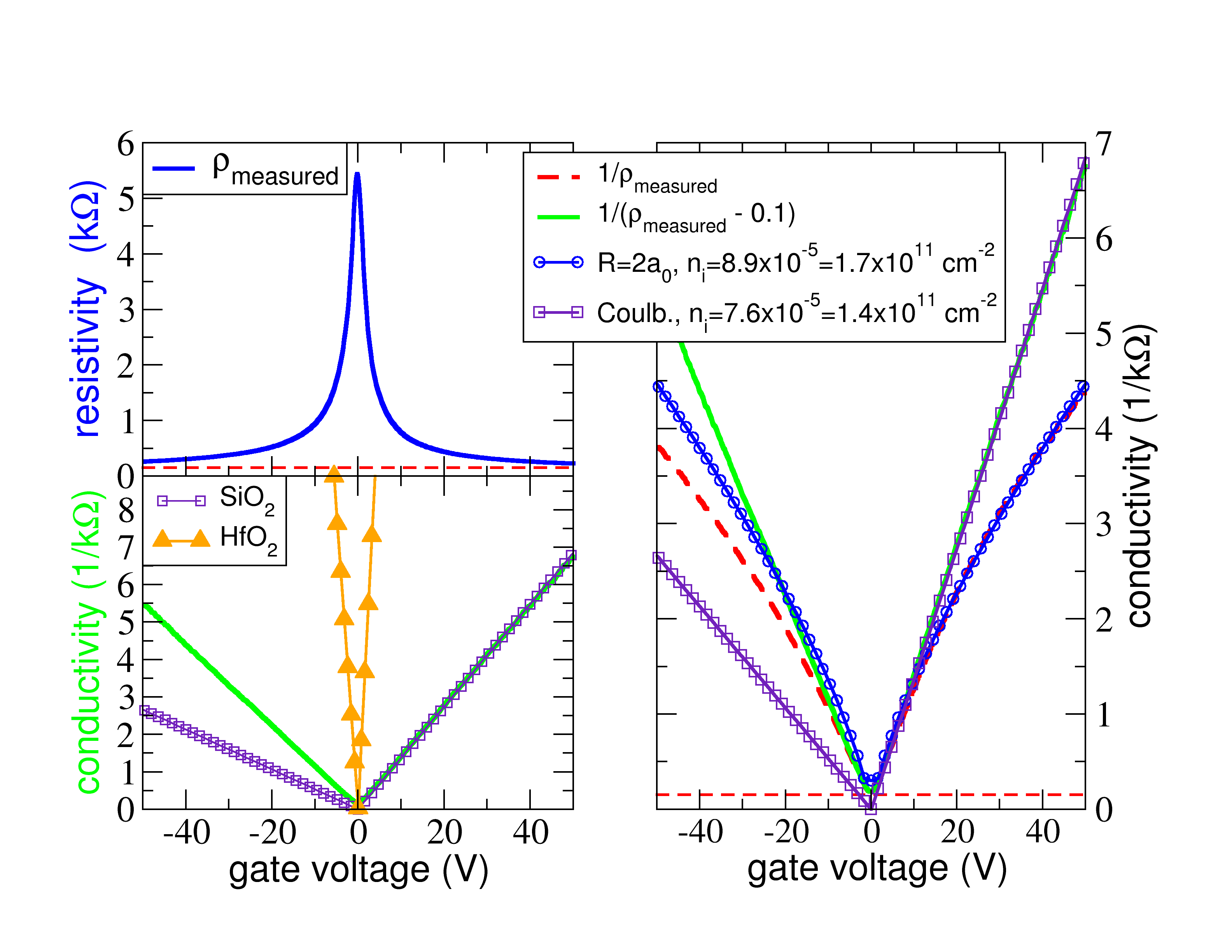}
\caption{(Color online)
Resistivity and conductivity of graphene on top of silicon oxide.
{\bf Top left panel:} Raw data of a measurement of the resistivity, $\rho_{\rm
measured}$, of an exfoliated graphene sheet
[data from S. V. Morozov {\it at al.} \cite{morozov}].
{\bf Right panel:}
Fit of the conductivity using Eq. (\ref{eq:sigma-midgap})
for the case where $\sigma_{\rm sub}=1/\rho_{\rm measured}$, and using
Eq. (\ref{eq:sigmaCoulomb}) for the case where
$\sigma_{\rm lin}=1/(\rho_{\rm measured}-\rho_S)$, with $\rho_S=100$ $\Omega$
[the value of $\rho_S=100$ $\Omega$ is that used by  S. V. Morozov {\it at al.}
\cite{morozov}].
{\bf Bottom left panel:} Data $\sigma_{\rm lin}=1/(\rho_{\rm measured}-\rho_S)$
fitted with the conductivity formula given by Eq. (\ref{eq:sigmaCoulomb})
(squares). For comparison, we give the
theoretical conductivity curve (triangles)
considering that the experiment had been done using HfO$_2$ as a dielectric.
This allows one to compare
 the modification of the numerical values
of $\sigma$ due to a substrate change.
(Data from S. V. Morozov {\it at al.} \cite{morozov},
courtesy of A. K. Geim.)}
\label{fig:chargescatterers}
\end{figure}

We have  now developed all the tools needed to
 perform the analysis of the data of Fig. \ref{fig:chargescatterers}.
We start by fitting the data using  the two models
presented above separately, that is Eqs. (\ref{eq:sigma-midgap}) and
(\ref{eq:sigmaCoulomb}).
We must stress that each of these two models have only one fitting parameter:
the concentration of impurities (in the model for resonant scattering, the
parameter
$R$ is fixed by the size of the primitive cell).
In the right panel of Fig. \ref{fig:chargescatterers} we plot
the data $\sigma_{\rm sub}=1/\rho_{\rm measured}$ and
$\sigma_{\rm lin}=1/(\rho_{\rm measured}-\rho_S)$
using the raw data $\rho_{\rm measured}$ (given in the left top panel of
the same figure). In the case of
$\sigma_{\rm sub}$, the data can be fitted using
Eq. (\ref{eq:sigma-midgap}) for mid-gap states. A perfect fit to both negative
and positive gate voltages is not possible, since by construction the
model developed in Sec. \ref{sec:mid-gapmodel} preserves electron-hole
symmetry; an improvement which does not preserve
electron-hole symmetry is easy to develop by considering a large finite
value (as opposed to an infinite value) for the effective
potential $g_{\rm eff}$ as discussed in Sec. \ref{sec:sourcesdisorder}
(see also \cite{stauberphonons,araujo});
 we show below that
charged scatterers can account for the loss of electron-hole symmetry of the
conductivity curves as well.

In the case
of the data computed as $\sigma_{\rm lin}$ we fit the positive gate voltage
region with Eq. (\ref{eq:sigmaCoulomb}) for charged scatterers.
Note that since $\Lambda(g)\ne\Lambda(-g)$ the computed conductivity
has no electron-hole symmetry, an effect seen in the
experiments \cite{fuhrer}. Nevertheless, although Eq. (\ref{eq:sigmaCoulomb})
does break electron-hole symmetry, the magnitude of the computed effect
 is far too strong, and therefore
 Eq. (\ref{eq:sigmaCoulomb})
 is not able to fit the data over the negative and positive range
of $V_g$, by assuming a single concentration of charged scatterers.
It is worth noting that the concentration of
impurities used to fit $\sigma_{\rm sub}$ and
$\sigma_{\rm lin}$
is essentially the same for both types of scatterers.
In the left bottom panel of Fig. \ref{fig:chargescatterers}
we depict the conductivity values (triangles),  had we performed the same
experiment
using HfO$_2$ as a dielectric.
To understand such a large change, we  look at Table
\ref{tab:sigmacoulomb},
 where we show  that $\Lambda(g)$ can be reduced by one order of magnitude
(positive $g$)
from SiO$_2$ to HfO$_2$,
leading to  the large increase in the conductivity shown in  Fig.
\ref{fig:chargescatterers}.

\begin{figure}[th]
\includegraphics*[width=7cm]{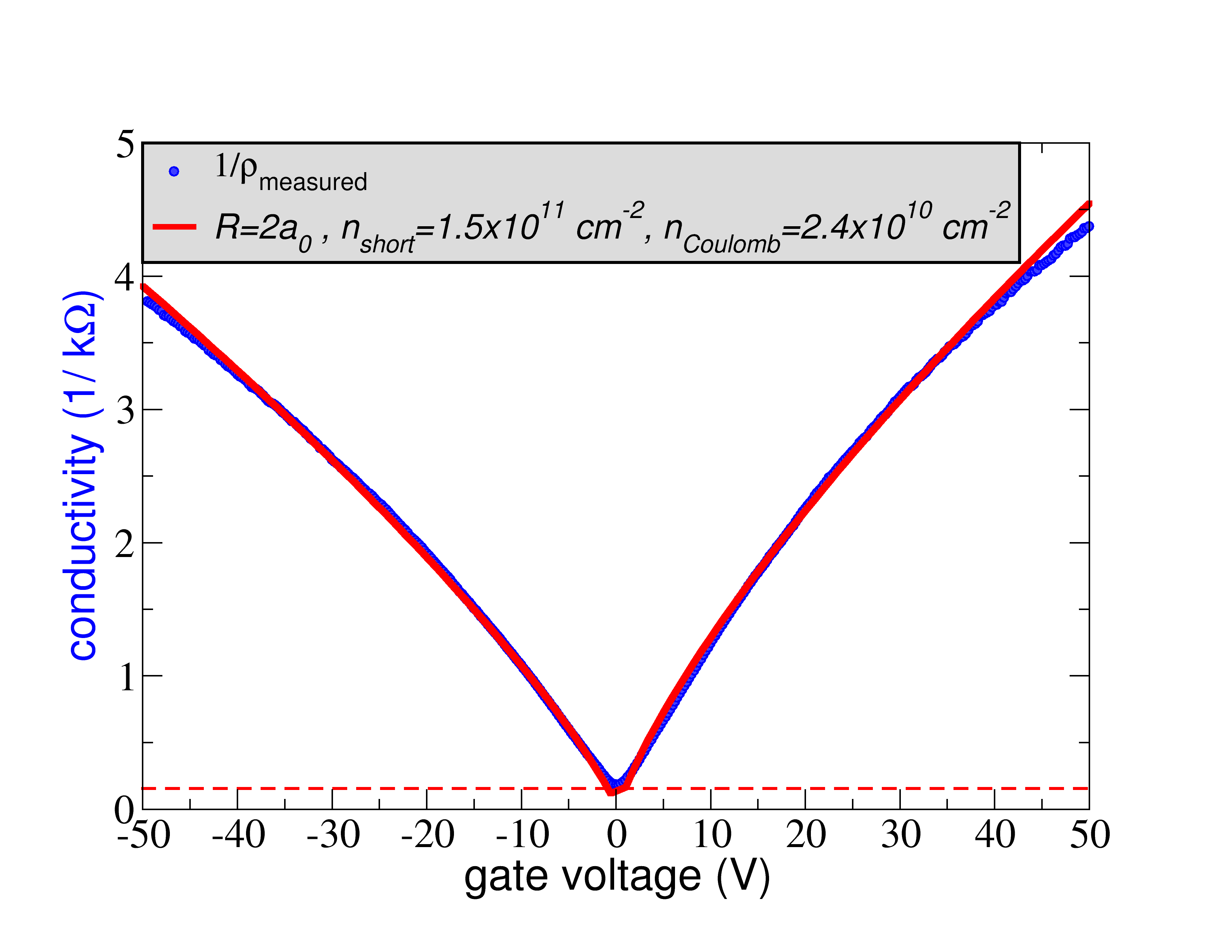}
\caption{(Color online)
The conductivity data
 $\sigma_{\rm sub}=1/\rho_{\rm measured}$, also plotted in
Fig. \ref{fig:chargescatterers}, but fitted using
Eq. (\ref{eq:sigmacombined}), which combines the effect of resonant and charged
scatterers.
In the legend, $n_{\rm short}$
 and $n_{\rm Coloumb}$ refer to the concentration of resonant and
charged scatterers, respectively.
The concentration of impurities is the only fitting parameter used in the
theory.
[Data from S. V. Morozov {\it at al.} \cite{morozov},
courtesy of A. K. Geim.)]}
\label{fig:sigmacombined}
\end{figure}

We now take into account, in a single model,  the effect of both strong
short-range and charged scatterers.
Computing
the conductivity as $\sigma_{\rm sub}=1/\rho$, such that (Matthiessen's rule)
\begin{equation}
\rho=\frac{1}{\sigma_{\rm short}}
+\frac{1}{\sigma_{\rm Coulomb}}\,,
\label{eq:sigmacombined}
\end{equation}
and with $\sigma_{\rm short}$ computed using Eq. (\ref{eq:sigma-midgap})
and $\sigma_{\rm Coulomb}$ determined from Eq. (\ref{eq:sigmaCoulomb}), we can
fit the data of
Fig. \ref{fig:chargescatterers} quite accurately, as shown
in Fig. \ref{fig:sigmacombined}.
In this figure,
the concentration of impurities leading to
mid-gap states used in the fit was
 $n_{\rm short}=1.5\times 10^{11}$ cm$^{-2}$ and that for charged scatterers
was $n_{\rm Coulomb}=2.4\times 10^{10}$ cm$^{-2}$, about seven times smaller
than $n_{\rm short}$ (these two concentrations are, essentially, the only two
fitting parameters
in the model).
The combined  contributions from resonant scatterers and charged impurities
allow one to fit $\sigma_{\rm sub}$  over the whole $V_g$ range, using  a single
value for the
impurity concentrations.

In conclusion,
we  made a thorough analysis of the role of resonant scatterers (which
give rise to mid-gap states) and charge scatterers
in the conductivity of graphene, and showed that a coherent picture emerges
from a scattering analysis of the transport
based on the exact calculation of the phase shifts of scattered chiral
Dirac fermions, as opposed to a calculation based on the first Born
approximation.

Finally, we  note that fine tuning details
coming from the dependence of the dielectric constant of graphene on the wave
vector
(the polarization contributions)
were not included in our simple model, except for the important effect of the
renormalization of
the fine structure constant, which corresponds to the large wave number limit.

Experiments will decide which scenario regarding the limiting source
of scattering in graphene actually prevails.

\subsection{Transport across a strained region: A way of generating a transport gap}
\label{sec:strain}

The main limiting factor of all graphene properties, in what concerns its
application
to nanoelectronics, is, most likely,  the lack of a true band gap, as opposed to
the
biased graphene bilayer \cite{McCann1,McCann2,castro}. This fact can, however,
be overcome
by creating a transport gap.

In nowadays nanotechnology, understanding the effect of strain on the
properties of  devices is an essential step toward the improvement
of their performance.
For example,
characterizing how strain can improve the properties of silicon-based devices is
a mainstream
research topic \cite{strainsilicon}.
As stated in Sec. \ref{sec:sourcesdisorder}, both ripples and wrinkles can 
act as scattering centers as they
effectively create
random strain in the material, leading to a modification of the hopping energy
$t$.
 In what follows, we show that strain in graphene
gives rise to a rich structure in the electronic and transport properties of the
material.

Being a 2D flexible membrane \cite{geimstiffness,kimneto},  stretching
(which in graphene can be as large as 20\%, being reversible) and
bending graphene in a controlled way is feasible
\cite{geimstrainraman,Ferralis:2008,Kim:2009}, with
consequences to the electronic \cite{vitorstrain} and
transport properties of the material
\cite{laustrain,vitorbreak,Fogler:2008,paconatphys}.
As we  see below, strain can be modeled by a fictitious
gauge field \cite{cortijoripples},
which  can then act as an effective magnetic field.
In some circumstances,  it was predicted that this effective magnetic field
can have an intensity as high as 10 T \cite{paconatphys}, leading
to a pseudo-magnetic quantum Hall effect. The presence of such
an odd quantum Hall effect can, in principle, be experimentally observed
using scanning tunneling microscopy, which is a direct measure of the
density of states, and, therefore, sensitive to the reorganization of the
spectrum due to the presence of the gauge field.
Such type of experiments have been performed \cite{Levy}
and found strain-induced pseudo-magnetic fields greater than 300 T.

In the case of suspended graphene \cite{laustrain} there are
two sources of strain. One is induced by the electric field produced
by the gate, which pulls the graphene membrane downwards. The
solution of the corresponding elasticity problem produces
an effective model where the effective vector potential is
constant \cite{Fogler:2008,vitorbreak}, precisely the model we
discuss below. The other source of strain depends on the thermal
properties of graphene.
Graphene's thermal expansion coefficient is anomalously large and negative
\cite{bao,geimthermal}, a feature which can be exploited
to induce
1D and 2D ripples (with a periodicity of about 300 nm)
possibly  leading to novel strain-based engineered  graphene devices
\cite{paconatphys,vitorbreak}.

\begin{figure}[ht]
\includegraphics*[width=7cm]{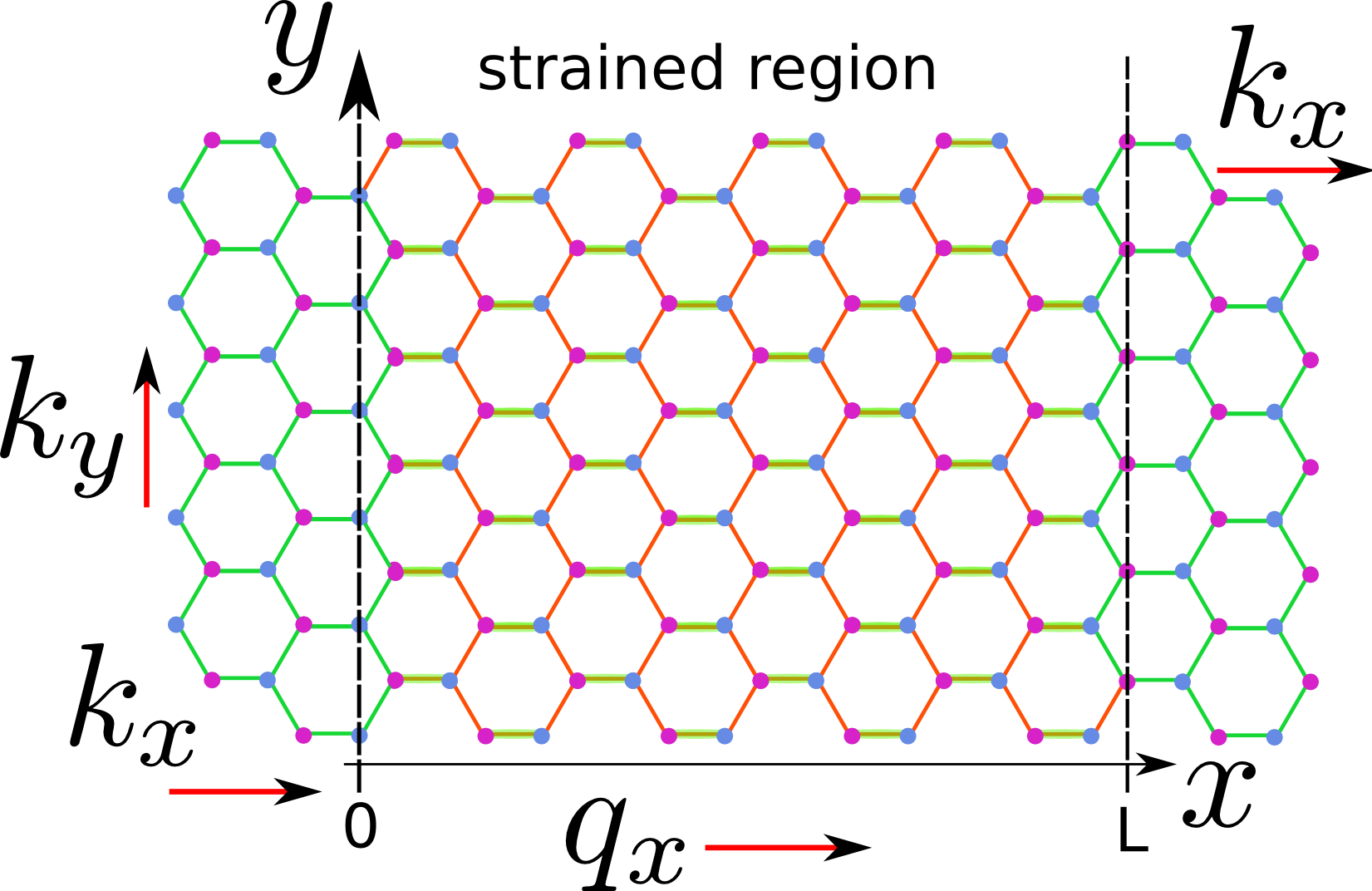}
\caption{(Color online)
Scheme of a device made of strained
graphene. The central region is the strained part. In the assumed model,
the effect of strain is to modify the hopping connecting a given
carbon atom to its three neighbors in such a way that two of the hoppings
are equal. Shown is the momentum of the electrons in the leads and in the strained
region.}
\label{fig:strain_tile}
\end{figure}

Consider a graphene-based device where the central part of the material,
of length $L$, is a graphene ribbon under strain, with armchair edges
oriented along the $x-$axis, as shown in Fig. \ref{fig:strain_tile}. The
strained part
is then connected to two pristine leads.
It is well understood that the effect of strain can be included
in the Dirac Hamiltonian in the form of a fictitious gauge
field \cite{rmp,paconatphys}. The emergence of  the fictitious
gauge field is simple to understand. We assume that the hopping
along the armchair edge is modified relatively to its
pristine value $t$ as $t\rightarrow t+\Delta t$ [a detailed study of how
strain
changes the value of the hopping was done using {\it ab initio} methods
\cite{ricardo}]. This adds a term to the
Hamiltonian, Eq. (\ref{eq:tbhamilt}), of the form
\begin{equation}
 \Delta t
\sum_{\bm R_n}(\vert A,\bm R_n\rangle \langle \bm R_n+a_0\bm u_x,B\vert
+H.c.)\,.
 \label{eq:gaugefield}
\end{equation}
Passing from the tight-binding description to the continuous model,
the contribution from  Eq. (\ref{eq:gaugefield}) has a finite value at a
given point in space. Since Eq. (\ref{eq:gaugefield}) couples the
sub-lattices $A$ and $B$ at the same point in space its contribution to
the effective Hamiltonian, Eq. (\ref{eq:dirac}), has the simple form
\cite{rmp}
\begin{equation}
\bm A(x,y)\sigma_y=-\theta(x)\theta(L-x)\frac{ \Delta t}{e v_F}\bm u_y
\sigma_y\,,
\label{eq:vector_A_strain}
\end{equation}
 implying that
the Dirac Hamiltonian maintains its original form,
but
with $\bm p$  replaced $\bm p\rightarrow (p_x,p_y+e\bm A_y)$. Clearly,
we have  a new Hamiltonian where the electrons now couple to a fictitious
vector potential $\bm A$ through the usual minimal coupling of
electrons to an electromagnetic field.

The question now is \cite{vitorbreak}: How are the transport properties of Dirac
electrons changed when transversing a region of strained graphene?
As usual, the answer to this question is obtained by
computing the transmission of the device by
matching the
wave functions from the left and right leads to those of the central region, at
the positions
$x=0$ and  $x=L$. From the matching conditions
we compute the total scattering matrix of the system \cite{quantumtransp},
relating
the incoming and outgoing waves. The scattering matrix $S$ is obtained easily
from the total transfer matrix of the structure $T_s$ using the same
formalism introduced in Sec. \ref{sec:ballistic}. In this case, the transfer
matrix
is given by
\begin{equation}
 T_s= \frac{1}{D}
\left[
\begin{array}{cc}
u&v\\
v^\ast & u^\ast
\end{array}
\right]
\left[
\begin{array}{cc}
u^\ast e^{-iLq_x}&-ve^{-iLq_x}\\
-v^\ast e^{iLq_x} & u e^{iLq_x}
\end{array}
\right]\,,
\label{eq:Tsmatrix}
\end{equation}
where $D=4\cos\theta\cos\tilde\theta$,
$u=e^{-i\theta}+e^{i\tilde\theta}$, $v=e^{-i\theta}-e^{-i\tilde\theta}$,
$\tan\theta=k_y/k_x$, $\tan\tilde\theta=(k_y-\delta)/q_x$, with
$k_y$, $k_x$, and $q_x$ the transverse momentum, the longitudinal momentum in
the leads, and
the longitudinal momentum in the device, respectively, and
$\delta=\Delta t/(v_F\hbar)$. Finally, the energy in the leads
has the form $\epsilon=\sqrt{k_x^2+k_y^2}$, and in the central
region $\epsilon=\sqrt{q_x^2+(k_y-\delta)^2}$. The last equation shows that
the effect of strain is to shift the position of the Dirac point in the
Brillouin zone, a crucial effect on the explanation of the following
results.
Equation (\ref{eq:Tsmatrix}) was derived for energies in the
continuum (no bound states). However, a fundamental property of the scattering
matrix (or the $T_s$ matrix for this purpose) is that bound states can also
be obtained from the form derived for  the scattering states,
by looking at the poles of the $S$ matrix.
Since the $S-$matrix is obtained from the inverse of
the $T_s$ matrix, its elements contain a factor which is the inverse of the
determinant of $T_s$.
 Additionally, the $S_{11}$
element of the $S-$matrix (in this problem the
$S-$matrix is a $2\times 2$ matrix, since we are working on the
propagating mode base) gives the amplitude
of transmission across the strained region, its value
being
\begin{equation}
S_{11}=\frac{4\cos\tilde\theta\cos\theta}
{\cos(Lq_x)
(\vert u\vert^2-\vert v\vert^2)-i\sin(Lq_x)(\vert u\vert^2-\vert v\vert^2)}\,.
\label{eq:S11}
\end{equation}
From Eq. (\ref{eq:S11}), we see that there are energies of perfect transmission,
when $Lq_x=n\pi$, with $n=1,2,\ldots$. On the other hand,
considering the case
$k_y=0$, corresponding to normal incidence on the boundary, we obtain for the
transmission
\begin{equation}
 T=\vert S_{11}\vert^2=\frac{\epsilon^2-\delta^2}{\epsilon^2-\delta^2\cos^2(Lq_x)}\,,
\label{eq_T_klein}
\end{equation}
which is a number smaller than one, meaning there is no Klein tunneling through
strained
graphene. At resonances, $Lq_x=n\pi$, the transmission (\ref{eq_T_klein}) is one,
and for energies
in the range $\delta/2<\epsilon<\delta$, it decreases 
exponentially with $L$, since $q_x$ becomes imaginary. 
In general, for  angles (between $\bm k$ and $y-$axis) 
satisfying the condition $\theta_f>\arccos(-1+\delta/\epsilon)$ the transmission, as computed from Eq. (\ref{eq:S11}), is
shown to be strongly suppressed for large $L$, an effect termed: 
transport gap
formation \cite{vitorbreak}. 
True energy gaps can also be
created in graphene by choosing an appropriate geometry of strain
\cite{paconatphys}.
The reason behind this transport gap mechanism is easy to understand given the
discussion
in Sec. \ref{sec:dispersion}; in the strained region the
electrons' wave function are no longer eigenstates of the helicity
operator $\hat h$. Therefore,  since the helicity is not a constant
of motion in this problem, Klein tunneling is lost and backscattering is
allowed.

\begin{figure}[th]
\includegraphics*[width=7cm]{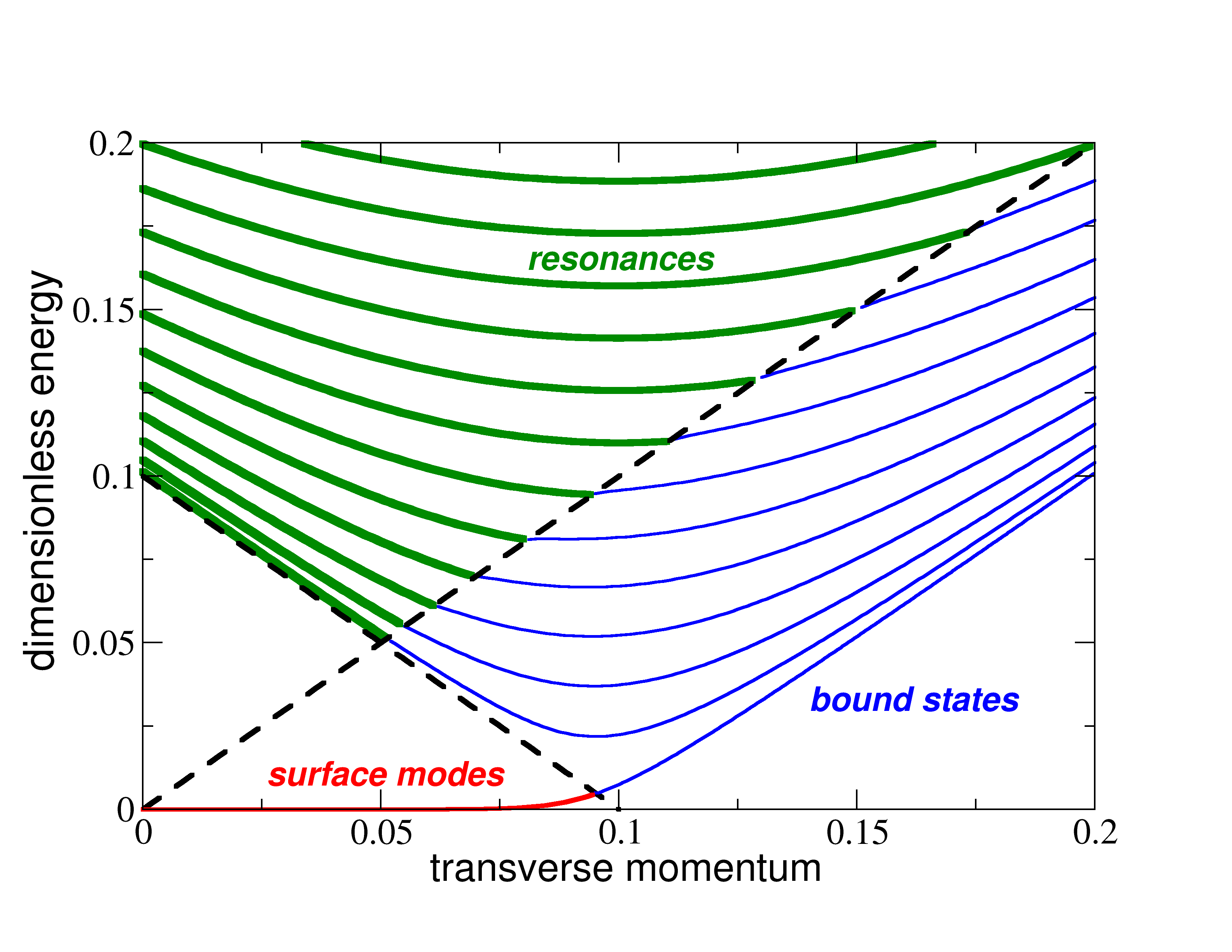}
\includegraphics*[width=7cm]{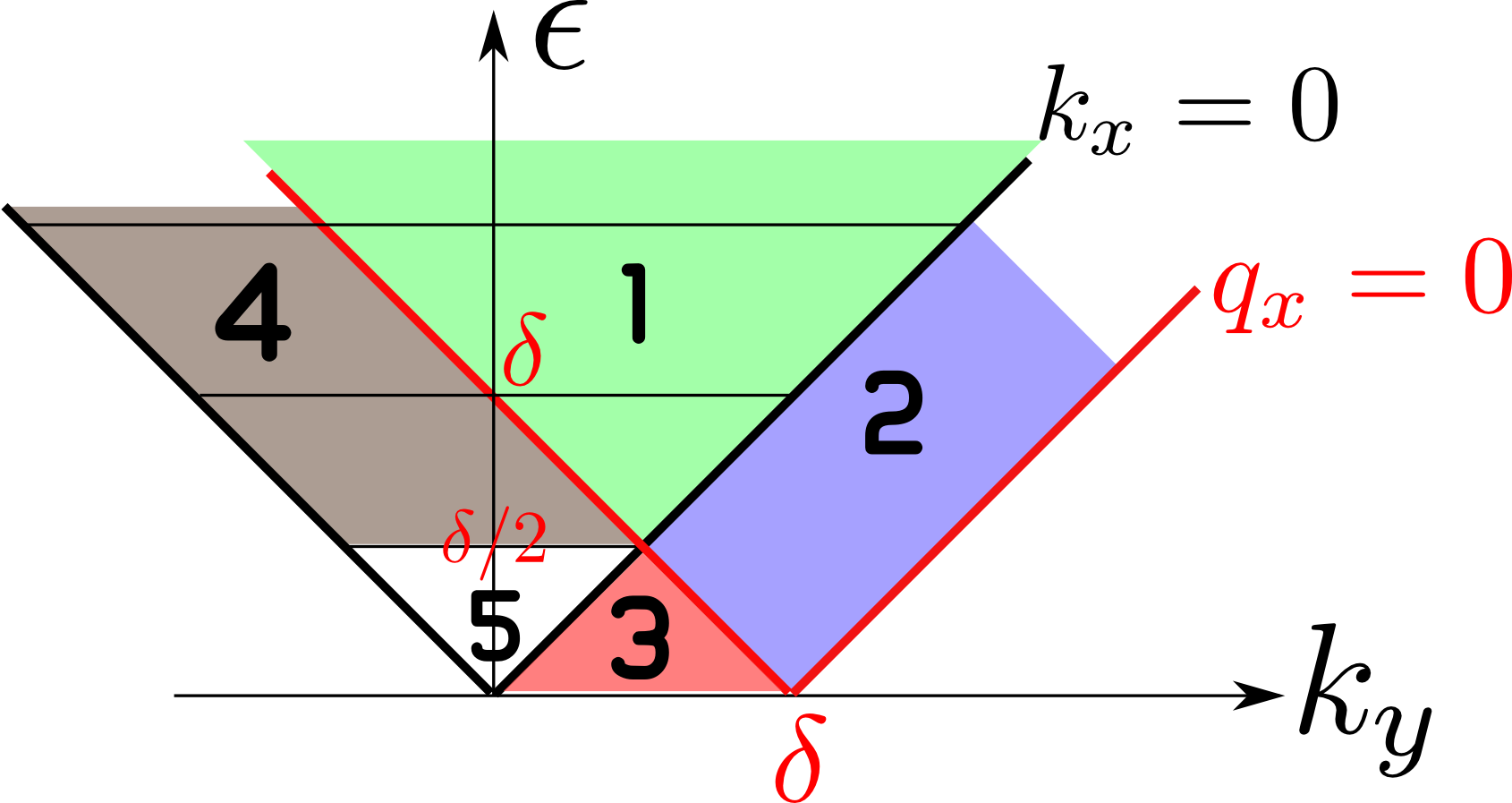}
\caption{(Color online)
Resonances, bound states, and surface states. 
{\bf Top panel:}
The parameters
are $L=100$ and $\delta=0.1$. In the resonances region, $\epsilon>k_y$,
the energy curves corresponds to the momenta $q_x=n\pi/L$.
The horizontal axis refer to the transverse momentum $k_y$,
and the vertical one to the dimensionless energy $\epsilon$.
{\bf Bottom panel:} Representation in the plane energy versus $k_y$
of the five types of states
appearing in this problem. The two Dirac cones, that of the contacts
and that of the strained region, are represented for 
$k_x=0$ and $q_x=0$, respectively. The off set of the apex of the two 
cones is $\delta$.
All states with $k_x\ne0$ and $q_x\ne0$ lie above the two respective cones.
}
\label{fig:strain_modes}
\end{figure}

As mentioned, the denominator of the $S_{11}$ element of the
$S-$matrix is all that it is needed to look for bound states in this system.
We can imagine two different types of bound states: those decaying
exponentially in the leads, but propagating inside the strained region and
those decaying exponentially in the three regions. The latter states
are edges states living at the boundaries between the leads and the strained
region. Looking only at the poles of  $S_{11}$ we can find that
both types of bound states exist \cite{vitorbreak}.
The richness of states in strained graphene is due to the breaking of the
chiral symmetry. In Fig. \ref{fig:strain_modes} we have represented
the energies at which the transmission is unity (called resonances), 
the energies of the bound states, and the energy of the edge or surface
states.
The different types of states are a result
of the different shifts of the two
Dirac cones over the Brillouin zone \cite{Montambeaux_2009,Hasegawa_2006b},
induced by the strain and governing the vector potential of Eq.
\ref{eq:vector_A_strain};
the two apexes are shift by $\delta$ due to strain, as seen above.

It was shown \cite{vitorbreak}that in the problem under 
study there are in total five types of states:
({\bf 1}) scattering states, ({\bf 2})
band states (states localized in the junction along the $x-$direction)
 propagating along the $y$
direction
({\bf 3}) localized states at the boundary of the junction, 
({\bf 4}) filtered states, that is, scattering states 
decaying exponentially inside the junction for certain values of the 
incoming angle $\theta_f$, and
({\bf 5}) and states such that the transmission occurs via
evanescent waves for any orientation of the incoming momentum -- it is said that
all states are filtered. 
The regions in the energy versus $k_y$ plane where 
these type of states appear are shown in the bottom panel of Fig. \ref{fig:strain_modes}.

In conclusion, the example discussed shows that there is  {\it plenty of room at
the
bottom} of strained graphene for a whole new sub-field of graphene research:
that of strain-based transport engineering or straintronics.

\section{Quantum corrections to the Drude conductivity}

Before discussing the quantum corrections to the Drude conductivity, we
introduce some key concepts on electronic transport.
Consider first the elementary transport theory
in a normal metal, where electrons have an effective mass $m^\ast$. The
velocity of
the electrons at the Fermi surface is given by $v_F=\hbar k_F/m^\ast$,
where $k_F$ is the Fermi wave number whose value depends on the density of
electrons.
Impurities in an otherwise perfect crystal
 occasionally deflect electrons from free propagation,
leading to the appearance of a mean free path $l$ --
the mean distance to the next collision.
Since the dominant contribution to transport comes from electrons having
velocity
$v_F$, we can introduce a phenomenological parameter, the relaxation time
$\tau$,
defined as $\tau v_F=\ell$;  the elementary theory of transport  then shows that
the electronic conductivity of the metal reads
\cite{ziman} (no spin or valley degeneracies included):
$\sigma_0=e^2n\tau/m^\ast$.
This is a purely classical result, known as Drude's formula, and which assumes
that,
after each collision, the electron looses memory of its previous linear momentum
state.
The calculation of $\tau$ is usually obtained from Fermi's golden rule.
The above description  makes sense when $\ell\gg\lambda_F=2\pi/k_F$ \cite{ziman}.
If we now repeat the same analysis for graphene, we obtain
\cite{peresBZ,stauberBZ,basko,sarna1,sarna2pnas}:
$
 \sigma_0=2e^2\tau v_Fk_F/h
$,
 where we have now included
the contributions of both spin and valley
degeneracies.

It is also possible to view the conductivity problem
as a random walk.
In this case, the conductivity is related to the
diffusion constant $D$ throught Einstein's relation \cite{schmid}:
$\sigma_0=e^2D\rho(\epsilon_F)$ (here again no spin or valley
degeneracies included), with $\rho(\epsilon_F)$ the density of states
per unit area, and
the units of $D$ are those of area per time,
in any spatial dimension. The diffusion constant, for order of magnitude
estimates, can be taken as $D\sim v_F\ell$.
\subsection{Weak localization in a normal metal}
\label{sec:weak}

Weak localization is a correction to the classical conductivity of
   a disordered metal due to quantum interference, and originates in
   the quantum mechanical superposition principle. Electrons propagating in
metals are
   subjected to a number of scattering mechanisms that give rise to a
   number of characteristic times.

The relaxation time $\tau$ due to
   elastic collisions with static impurities is assumed to be the
   smallest scattering time and describes a reversible process. Other
   scattering mechanisms are irreversible in nature and lead to either
   the loss of phase coherence or energy relaxation; for instance,
   those caused by electron-electron and electron-phonon interactions
   (excluding interactions with magnetic impurities). Contrary
   to $\tau$, the phase relaxation (or dephasing) time $\tau_\varphi$ is
   temperature dependent and at low temperatures is mainly due to
   electron-electron interactions. In the presence of a magnetic
   field, a new time scale $\tau_B$ appears,
which  is of  the order of
$\tau_B \sim \ell^2_B/D$, by simple dimensional analysis arguments,
and where $\ell_B^2=\hbar/eB$ is the magnetic length.
 We now proceed to the
discussion of the quantum interference effects
using an intuitive approach \cite{abrikosov,houten},
rather than a formal one, as seems appropriate in the context of this
Colloquium.
\begin{figure}[th]
\includegraphics*[width=8cm]{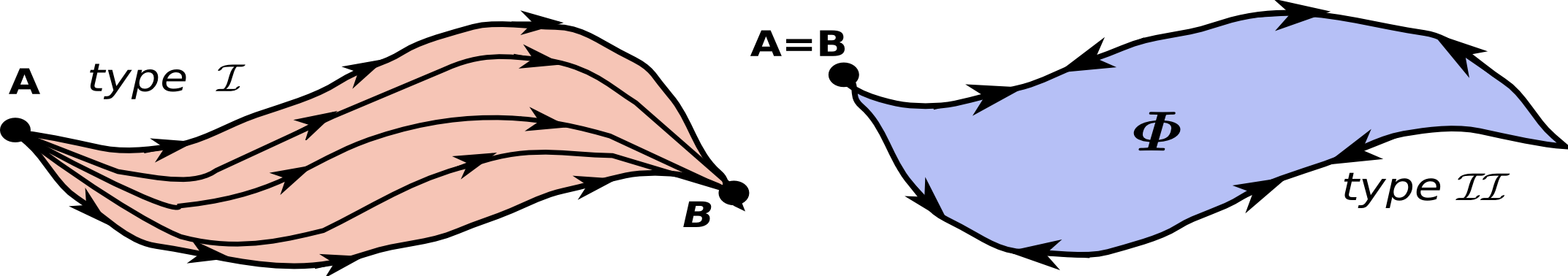}
\caption{(Color online)
Pictorial representation for two types of scattering processes, whose physical
interpretation
is given in the text. The presence a finite magnetic field is represented by
the flux $\Phi$ piercing the
area defined by the two
time reversed trajectories.}
\label{fig:weak}
\end{figure}

Imagine an electron traveling from position $A$ to $B$, as
shown in Fig. \ref{fig:weak}, and we denote by $a_{i}e^{i\phi_{i}}$ the probability
amplitude for the electron to travel from $A$ to $B$, along trajectory
$i$. Since there are many indistinguishable trajectories, the total
probability of traveling from $A$ to $B$ is
\begin{equation}
P_{I}(A\to B)=\left|\sum_{i}a_{i}^{i\phi_{i}}\right|^{2}\,.
\end{equation}
We now show that quantum interference effects are much more
important for what we call trajectories of type II, in Fig. \ref{fig:weak}, in
which
the initial and final points coincide ($A=B$).

For type I trajectories ($A\neq B$), we have
\begin{equation}
P_{I}(A\to B)=\sum_{i}\left|a_{i}\right|^{2}+
\sum_{i\neq j}a_{i}a_{j}e^{-i\left(\phi_{i}-\phi_{j}\right)}.
\label{eq:eq1}
\end{equation}
Since the phases for different trajectories of type I are uncorrelated,
we  assume that the second term averages to zero, leaving us with
the classical result, in which the probability to go from $A$ to
$B$ is just the sum of the probabilities over all possible trajectories, that is,
\begin{equation}
P_{I}(A\to B)=\sum_{i}\left|a_{i}\right|^{2}=P_{I}^{(cl)}(A\to B)\,.
\end{equation}
However, for trajectories of type II (the same initial and final points),
in the presence of time reversal symmetry, the situation is quite
different. In fact, time reversed trajectories (going round the loop
in clockwise and anti-clockwise fashion), contribute to the sum of
Eq. (\ref{eq:eq1}) with the same amplitude,  in both modulus and \emph{phase}. As
a result, in the interference term,
\begin{equation}
\sum_{i\neq j}a_{i}a_{j}e^{-i\left(\phi_{i}-\phi_{j}\right)}\,,
\label{eq:eq2}
\end{equation}
when $i$ and $j$ denote time reversed trajectories, the phases cancel,
even before any averaging; this term gives a contribution exactly
equal to the first one, since for every trajectory there is a time reversed
pair. This then amounts to a probability
\begin{equation}
P_{II}(A\to A)=2\sum_{i}\left|a_{i}\right|^{2}=2P_{I}^{(cl)}(A\to A) \,.
\label{probII}
\end{equation}
This effect of quantum interference therefore enhances the probability
of return, relative to the classical result, decreasing diffusion
and, therefore, the conductivity \cite{abrikosov,houten}; in other
words we have
\begin{equation}
\sigma_{wl}-\sigma_0 < 0 \,,
\label{sigmaWL}
\end{equation}
where $\sigma_{wl}$ stands for the conductivity of the metal considering
the enhanced backscattering effect, due to quantum interference. It is the
reduction in $\sigma_{wl}$ relatively to $\sigma_0$ that is known as
weak localization.

In the presence of a magnetic field $\bm B$ the relative phase of the
electron's wave function, associated with the two time reversed trajectories of
type
II, has the value $\delta\phi=4\pi\Phi/\phi_0$ as given
by the Aharanov-Bohm effect,  where $\Phi$ is the magnetic
flux piercing the area defined by the closed trajectory. Therefore, applying a
magnetic field to the system suppresses the interference effect
(because it changes the relative phase to a
non-zero value)
given by Eq. (\ref{probII}), and the low-temperature
conductivity of the metal increases when the field
is turned on; or, in other words, we have
\begin{equation}
\frac{\delta\sigma(B)}{\sigma_{wl}}
\equiv\frac{\sigma_{wl}(B)-\sigma_{wl}}{\sigma_{wl}}>0\,.
\label{eq:sigmaWLB}
\end{equation}
Using Eq. (\ref{eq:sigmaWLB})  we  obtain experimental
evidence of
weak localization effects in the conductivity of a disordered metal.

If the spin-orbit interaction \cite{larkin}
can be ignored and there are no magnetic
impurities in the metal, the  times $\tau$, $\tau_\varphi$, and $\tau_B$
are the only relevant time scales, and they
control the behavior of the low temperature conductivity.

We now extend the previous analysis to graphene. Again,
as expected, the chiral nature of the electrons (or, equivalently,
their non-trivial Berry's phase) will play a major role.
As before, we  keep the discussion as elementary as possible.

\subsection{Weak localization in graphene}
\label{sec:weakgraph}

In graphene, the quasi-exact conservation of the chirality
and the existence of two valleys have  profound effects in the low-temperature
conductivity of the material.
Below, we present the general picture of the quantum corrections in graphene,
referring the interested reader to the  literature for the subtleties
appearing under a
detailed analysis of this problem 
\cite{morpurgoWL,mccann,andoWL,aleiner,eduardoreview}.

As shown in Sec. \ref{sec:weak}, in a normal metal the only
elastic time is the relaxation time $\tau$. In graphene
the situation is more complex. In order to understand
the complexity of the situation,  consider two different matrix elements
of a
potential created by a given impurity. We assume the potential
to have the form
\begin{equation}
V(\bm r) = \frac{u}{r_0^2\pi}e^{-r^2/r_0^2}\,.
\end{equation}
The range of the potential depends on the value of $r_0$: the larger
$r_0$, the larger the range.
The effect of this potential on  electrons within the same valley
(denoted intra-valley scattering)
is given by the matrix element
\begin{equation}
\langle \psi_+(\bm k')\vert
V(\bm r)\vert\psi_+(\bm k)\rangle
=\frac{u}{8A_c}f(\bm k,\bm k')
e^{-q^2r^2_0/4}\,,
\label{eq:ampintra}
\end{equation}
with $f(\bm k,\bm k')=\cos[\theta(\bm k)/2-\theta(\bm k')/2]$
and $q=\vert \bm k-\bm k'\vert$. If we take $\epsilon_F=0.5$ eV one obtains
$q\sim 0.2/a$,
with $a=\sqrt 3 a_0$.
The function $f(\bm k,\bm k')$ shows that the scattering
is not isotropic in momentum space, a consequence of the
chiral nature of  electrons in graphene. From
 $f(\bm k,\bm k')$, we also see that the scattering amplitude
for backscattering, $f(\bm k,-\bm k)$,
is zero, the fingerprint of Klein tunneling (recall Fig. \ref{fig:scattering})
for massless Dirac electrons
\cite{beenakkrmp}.

If the potential also couples electronic momentum states from $\bm K$
and $\bm K'$ valleys (denoted inter-valley scattering) the
matrix element, using wave functions from different valleys, reads
\begin{equation}
\langle \psi_+(\bm k')\vert
V(\bm r)\vert\psi_+(\bm k)\rangle
=\frac{u}{8A_c}g(\bm k,\bm k')
e^{-Q^2r^2_0/4}\,,
\label{eq:ampinter}
\end{equation}
with $g(\bm k,\bm k')=i\sin[\theta(\bm k)/2-\theta(\bm k')/2]$
and $Q\simeq\vert \bm K-\bm K'\vert=4\pi/(3a)$. In this case,
backscattering is permitted, $g(\bm k,-\bm k)\ne 0$, since
 scattering couples states in the $\bm K$ and $\bm K'$
valleys, which have opposite chirality (recall Fig. \ref{fig:scattering}).

Equation (\ref{eq:ampinter}) shows that only for very 
short-range potentials, $r_0\lesssim a$,  does intervalley scattering
have a significant amplitude. For long range potentials, only intra-valley
scattering plays a role.

From the above discussion it follows
that, in the case of graphene, we need to
 define several
elastic scattering times \cite{morpurgoWL,mccann,aleiner}:
\begin{enumerate}
\item $\tau_{iv}$, representing inter-valley scattering, whose scatterers
are very short-range potentials with range  $r_0\lesssim a$, such as
some types of adatoms, adsorbed hydrocarbons, or
vacancies.
\item $\tau_s$, representing intra-valley scattering, whose
scatterers are long range potentials, such as ripples, dislocations, and
charged scatterers.
\item $\tau_w$,  representing also another contribution to intra-valley
scattering.
This scattering time
has its origin in the fact that chirality is not an exact symmetry
of Dirac fermions in graphene (due to trigonal warping effects),
therefore allowing for some
amount of backscattering within the same valley.
The importance of this scattering time grows as the
Fermi energy increases.
\end{enumerate}
As long as the scattering potentials are long range, inter-valley scattering
is negligible and back-scattering in graphene is absent, except from a small
contribution from $\tau_w$. The effect just described is, at the more
fundamental
level, a consequence of the
Berry's phase (of $\pi$) acquired by massless Dirac electrons when
they perform a closed chiral orbit \cite{berry,mikitik}. The Berry's phase
transforms the constructive interference, we described above for normal metals,
into a destructive one, leading to  weak anti-localization.
Under these circumstances, the interference
effect seen in Sec. \ref{sec:weak} for a normal metal cannot exist
in graphene, forward scattering is enhanced
and one should expect weak anti-localization effects to manifest
themselves, that is, if $\tau_{iv}\gg\tau_\varphi$ we have (ignoring $\tau_w$)
\cite{andoWL,mccann}
\begin{equation}
\sigma_{awl}-\sigma_0>0\,,
\label{sigmaAWLg}
\end{equation}
where $\sigma_{awl}$ represents the enhancement (anti-localizing effect) of the
conductivity over
Drude's result $\sigma_0$ due to Klein tunneling.
If short range scatterers are present, then inter-valley scattering
plays a role, and since electrons in the ${\bm K}$ and ${\bm K'}$ valleys
have opposite chirality,
back-scattering is present [see the function
$g(\bm k,\bm k')$], and we  expect weak localization effects,
according to Eq. \ref{sigmaWL};
detailed calculations confirmed this picture \cite{andoWL,mccann}.
When the effect of a magnetic field is included, the general rigorous expression
for the
weak localization corrections was derived by two groups independently
\cite{mccann,aleiner},
and  $\delta\sigma(B)$ can  be either positive or negative depending on the
relative values of the different scattering times, including $\tau_B$.
For a comprehensive discussion of the interplay between the different scattering
times and the quantum corrections to the conductivity, the interested
reader is referred
to the technical literature \cite{aleiner,mccann,mccannII}.
\begin{figure}[ht]
\includegraphics*[width=8cm]{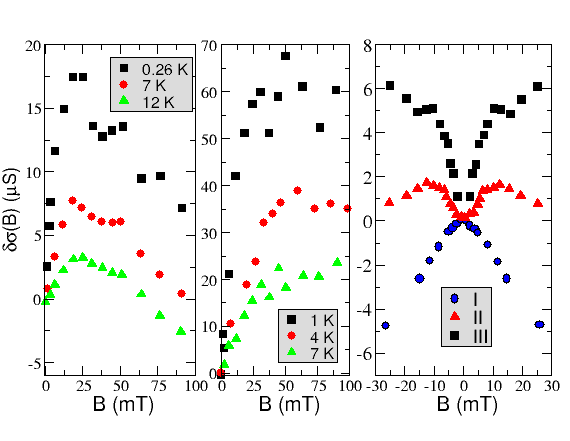}
\caption{(Color online)
The quantity $\delta\sigma(B)$ is defined
as in Eq. (\ref{eq:sigmaWLB}). {\bf Left:} 
Weak anti-localization behavior (the  gate voltage used was
$V_g\lesssim 1$ V, corresponding to
an electron density $n\lesssim 7\times 10^{10}$ cm$^{-2}$).
{\bf Center:}
Weak localization behavior (the  gate voltage used was
$V_g= 11$ V, corresponding to
an electron density $n\simeq 8\times 10^{11}$ cm$^{-2}$).
{\bf Right:} Dependence of $\delta\sigma(B)$ on the
electronic density (it grows from I to III), at $T=27$ K.
A fit of the data must  use
the rigorous formula derived in the literature
\cite{mccann,aleiner}.
[Data from F. V. Tikhonenko {\it et al.}
\cite{savchenko,Tikhonenko2009}.]}
\label{fig:wl}
\end{figure}

In the left panel
of Fig. \ref{fig:wl} we show, for small values of the magnetic field,
a weak localization dip ($\delta\sigma (B)$ grows) followed by, above a certain
field value $B^\ast$,   weak anti-localization
behavior of the conductivity, since  $\delta\sigma (B)$
starts to decrease upon increasing the
magnetic field over $B^\ast$ (no saturation of
 $\delta\sigma (B)$ is measured upon increasing $B$, as in the weak localization
case).
  In the central panel of Fig. \ref{fig:wl}, we show
weak localization behavior in graphene, since the corrections $\delta\sigma (B)$
never
decrease upon increasing the magnetic field, and tend to saturation. Since the
electronic
density in the central panel of Fig. \ref{fig:wl} is few times larger than that
in the left one,
it seems that the effect of short-range scatterers is more effective at higher
densities
\cite{Tikhonenko2009}; at lower densities long-range scatterers dominate.
Indeed, in the right panel of Fig. \ref{fig:wl}, we clearly see a crossover from
weak anti-localization to weak localization as the electronic density
increases from I to III, at a temperature of 27 K; considering important
screening effects of charged impurities at large electronic densities, such a
result
sounds reasonable. It is a remarkable experimental
fact that quantum interference effects  just discussed can be observed
in graphene
at temperatures as high as $\sim$200 K \cite{Tikhonenko2009}.

As stated, we expect that in graphene the presence
of  different kinds of defects (in different concentrations) will control
whether  weak localization
or weak anti-localization is observed. This depends on the relative
value of the different elastic times introduced above and on the electron
density.
A detailed
analysis of this point is essential for a correct interpretation of the
data, and has been done  in great detail
\cite{mccann,morpurgoWL}.
The numerical values for the different scattering times can be obtained from the
experimental data \cite{savchenko}.

Finally, we note that the observation of quantum corrections to the
conductivity in graphene
seems to depend on the details of the fabrication process, which determines
the amount
of rippling introduced in the system \cite{geimWL}.
Routes for suppression  of weak (anti-)localization effects
have been considered  \cite{khveshchenkoWL} and this effect was
experimentally observed as well \cite{savchenko,geimWL}.
As in the case of strain
discussed in Sec. \ref{sec:strain}, ripples are equivalent to effective
gauge fields which break time reversal, leading to the suppression
of weak localization effects, within each valley. In the system as a whole
(both valleys considered)
the full time reversal symmetry is preserved.

\section{The optical conductivity of graphene in the infrared to visible range of the spectrum}
\label{sec:optical}

In the ensuing sections we discuss the calculation of the percentage
of light transmitted by a graphene membrane, when light shines from behind.
This property is controlled by the optical conductivity $\sigma(\omega)$
of the material. We  analyze how and why the experimental behavior
of $\sigma(\omega)$ deviates from the predictions of the independent electron
model.

\subsection{Graphene as a transparent membrane}
\label{sec:transmittance}

\begin{figure}[ht]
\includegraphics*[width=7cm]{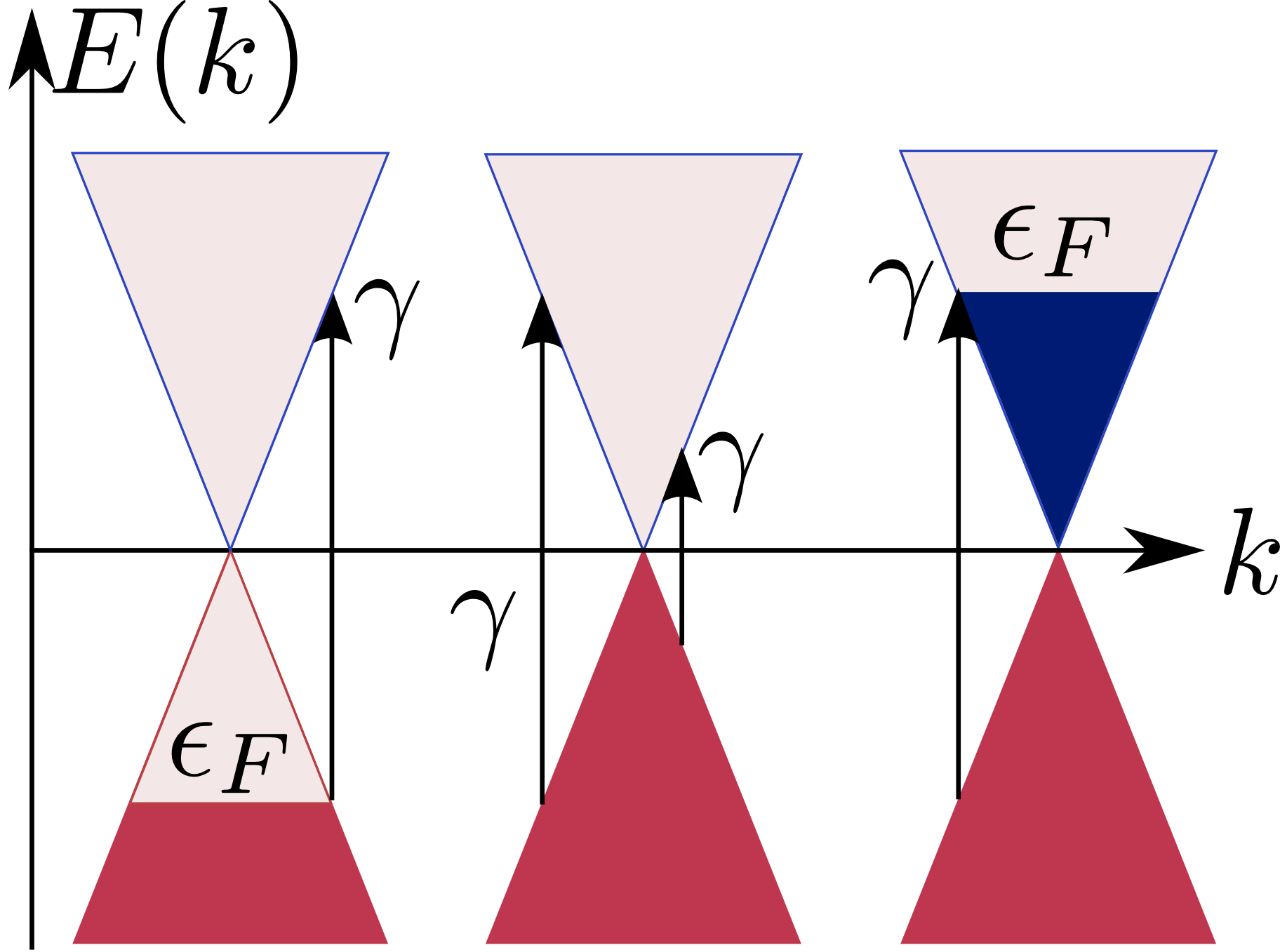}
\caption{(Color online)
Pictorial description of the optical excitation of electrons
in graphene. The absorption of a photon can only induce vertical inter-band
transitions. From left to right
we have graphene doped with holes, neutral, and doped with electrons.}
\label{fig:optical_pictorial}
\end{figure}

The calculation of light absorption by a given material is equivalent
to the calculation of the optical conductivity. In general, such a calculation
proceeds using  Kubo's formula. In the case of graphene, it is
possible to
use Fermi's golden rule to obtain directly the fraction of absorbed light,
 which turns
out to be a much simpler calculation than computing  the
optical conductivity first \cite{Geimellipsometry}. The
central quantity to be computed is the transition rate of electrons
excited from the valence band to the conduction one, as shown
in Fig. \ref{fig:optical_pictorial}.

\begin{figure}[th]
\includegraphics*[width=5.5cm]{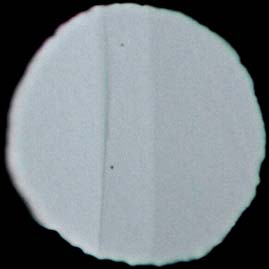}
\caption{(Color online)
 An optical image of an  aperture partially covered
with graphene and its bilayer (from left to right: air/graphene/bilayer),
taken in a light transmission experiment
(courtesy of A. K. Geim).}
\label{fig:opticalgeim}
\end{figure}

In the presence of a vector potential $\bm A$ the Dirac Hamiltonian has the
form
\begin{equation}
H_{\bm K}=v_F\bm \sigma\cdot (\bm p + e\bm A)\,.
\end{equation}
We  represent the electric field as $\bm E=-\partial \bm A/\partial t$
and choose the polarization of the field along the $x-$axis:
$\bm A=\hat x A_0(e^{i\omega t}+e^{-i\omega t})/2$.
The term $v_F\bm \sigma\cdot e\bm A$ will be taken as perturbation, and in the
spirit of
time dependent perturbation theory, only the exponential with negative exponent
is
taken.
The transitions induced by light absorption are now controlled by the
$\sigma_x$ matrix.
Clearly the matrix element $\langle \psi_\lambda\vert\sigma_x\vert
\psi_\lambda \rangle$ cannot contribute to the conductivity, since light cannot
induce
transitions within the same band, among states of equal momentum.
The only non-vanishing contributing
matrix element
is therefore $\langle \psi_1\vert\sigma_x\vert
\psi_{-1} \rangle=-\frac{i}{2} v_FeA_0\sin\theta(\bm k)$. The
transition rate is then given by Fermi's golden rule:
\begin{equation}
W_{1,-1}(\bm k)=\frac{2\pi}{4\hbar} v_F^2e^2A_0^2
\sin^2\theta(\bm k)\delta(2v_Fk\hbar-\omega\hbar)\,.
\label{eq:matrix_rate}
\end{equation}
The Dirac delta function in Eq. (\ref{eq:matrix_rate}) enforces the
condition that only electrons with energy $\omega/2$ can be excited to the
conduction band. The transitions we are referring to are shown in
Fig. \ref{fig:optical_pictorial}. To obtain the   contribution from all states
we have
to integrate over the momentum and multiply the result by four (two for spin
times two
for valley).  The calculations are
elementary and the result for the total transition rate per unit area is
\begin{equation}
 \frac{1}{\tau}=\frac{e^2A_0^2\omega}{8\hbar^2}\,.
\end{equation}
If light of frequency $\omega$ is shining upon a unit area of graphene,
the amount of absorbed power per unit area is $W_a=\hbar\omega/\tau$.
The energy flux impinging on graphene is given by
$W_i=c\epsilon_0E_0^2/2$, with $E_0=A_0\omega$. Therefore the
fraction of transmitted light is \cite{nair}
\begin{equation}
T=1-\frac{W_a}{W_i}=1-\pi\alpha\simeq 0.977\,,
\label{eq:transmission}
\end{equation}
with $\alpha=e^2/(4\pi\epsilon_0\hbar c)$ the fine structure constant. The
absorption of light
is therefore independent of frequency and given only
by universal constants.
The high transmittance of graphene is shown Fig. \ref{fig:opticalgeim}; 
it is remarkable that a one atom
thick membrane
can be seen by the naked eye.

It follows from the previous analysis that the transmission at finite
doping is given by
\begin{equation}
T(\epsilon_F)\approx(1-\pi\alpha)\theta(\omega-2\epsilon_F)\,,
\label{eq:transmission_finite_mu}
\end{equation}
where
the Heaviside step function takes into account that absorption can only occur
for frequencies larger than twice the Fermi energy, due to Pauli's principle.
The result given by Eq. (\ref{eq:transmission})  is identical to the one given
by a
rigorous calculation based on Kubo's formula
\cite{nmrPRB06,abergel,gusyninIJMPB,peresIJMPB,kuzmenko}.
The reason why the perturbative calculation works so well is because
the final answer is controlled by the small dimensionless parameter
$\alpha$. The reason why the transmission is controlled by the fine
structure constant originates in the chiral nature of the
electrons in graphene, a result extensible to few-layers graphene
\cite{min}.

It is now a simple matter to include corrections to the linear spectrum of
graphene in the formalism (the conical nature of the spectrum is
valid  for energies of the order of
$\lesssim 1$ eV). The addition of a next-nearest neighbor hopping term
can also be included and treated within this formalism. The case
of a next-nearest neighbor hopping is actually trivial, being
proportional to the identity matrix its contribution is zero.
In fact, it is proven in general \cite{staubergeim} that the
contribution of the next-nearest neighbor hopping only enters
in the final result as a renormalization of the energy spectrum.

As stated, the transmittance of light through graphene can be computed
from a previous knowledge of the optical conductivity of the material.
The  transmittance is calculated from the
solution of Fresnel's equations, reading
\cite{staubergeim,blake,abergel,pedersen}
\begin{equation}
T=\vert 1 +
\sigma(\omega)/(2c\epsilon_0)\vert^{-2} \,,
\label{eq:fresnel}
\end{equation}
where
$\sigma(\omega)$ is the
the optical conductivity
\cite{nmrPRB06,staubergeim,falkov1,falkov2,gusynin3,gusyninIJMPB,carbotte} of
graphene,
given, at zero temperature and within the independent electron approximation, by
($\epsilon_F>0$)
\begin{equation}
\sigma(\omega)=\sigma_0\theta(\omega\hbar-2\epsilon_F)+i\sigma_0\frac{
4\epsilon_F}{\pi\omega\hbar}
-i\frac{\sigma_0}{\pi}\ln\frac{\vert \hbar\omega+2\epsilon_F\vert}{\vert
\hbar\omega-2\epsilon_F\vert}\,.
\label{eq:sigoptical}
\end{equation}
The quantity $\sigma_0=\pi e^2/(2h)$ is termed the ac universal conductivity of
graphene.
Inserting Eq. (\ref{eq:sigoptical}) into Eq. (\ref{eq:fresnel}) and taking
$\epsilon_F=0$ we obtain
the result of Eq. (\ref{eq:transmission}). 
Working the other way around, using Eq. (\ref{eq:transmission_finite_mu})
in Eq. (\ref{eq:fresnel}), we obtain for neutral graphene
$\sigma(\omega)=\sigma_0$, in accordance with Eq. (\ref{eq:sigoptical}).
The fact that
for neutral graphene
$\sigma(\omega)$ is given in terms of universal constants only,
with no reference to any of the material parameters,
is a rare result in condensed matter physics.
If the intensity of light impinging on graphene is {\it large},
then non-linear corrections to $T$
start to play a role, which  is expected to lead to an increase of
the transmittance \cite{Michenko,Rosensteinnon-nonlinear}.
The non-linear optical susceptibility coefficients of graphene have recently been
measured using four-wave mixing  \cite{Nonlinearsusceptibility}.

Due to its high transmittance,  graphene can be used as
transparent conductive electrodes  in solar cells and
liquid crystal devices (LCD) \cite{solar,geimliquid,rolltoroll}.
\subsection{The optical conductivity of neutral graphene}
\label{sec:sigma_neutral}

We now discuss the experimental results for the optical conductivity
of neutral  graphene and how those measurements deviate from the
independent electron model  presented above.

\begin{figure}[th]
\includegraphics*[width=7cm,angle=0]{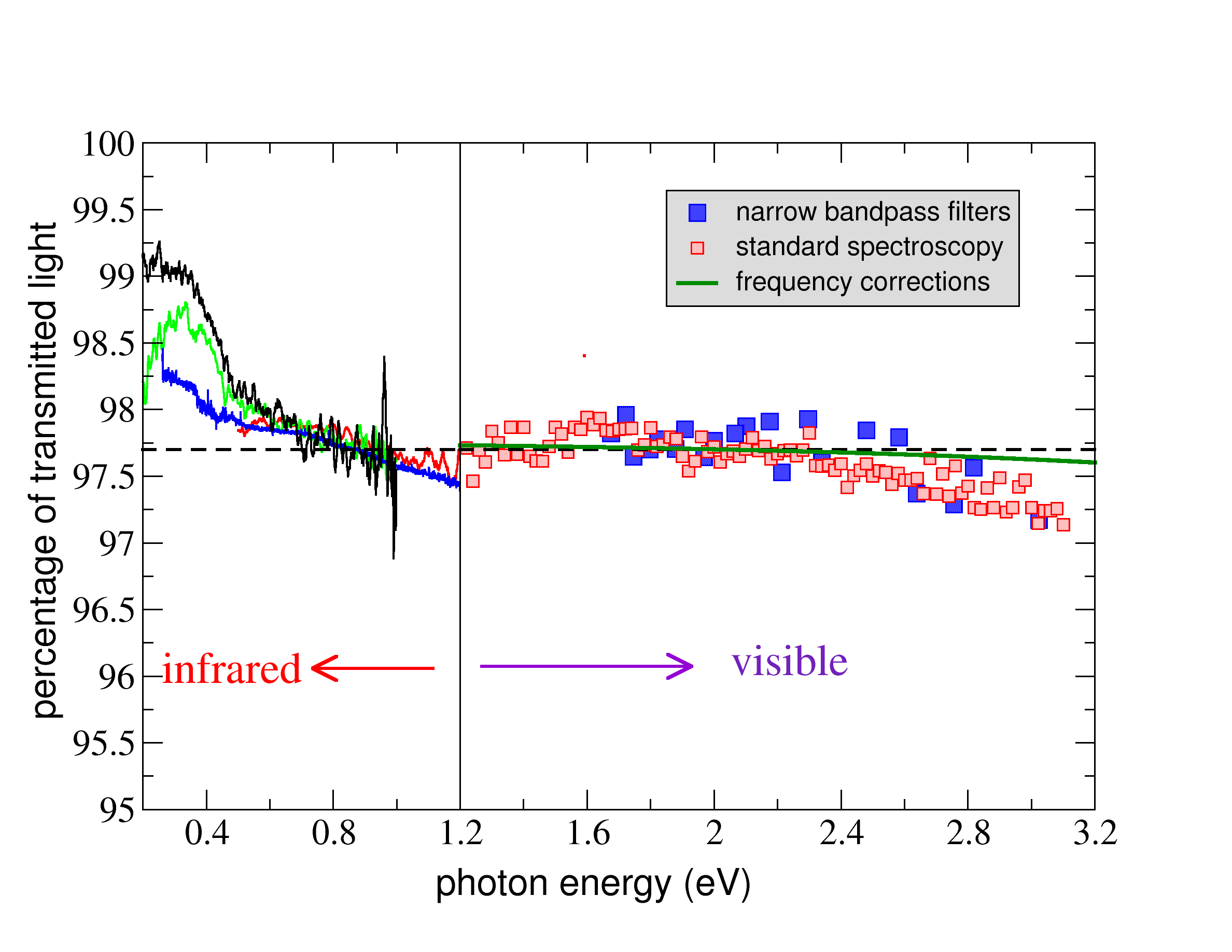}
\caption{(Color online)
Transmittance of graphene.
For photon energies below 1.2 eV, we plot data from
reflectance measurements taken by K. Fai Mak {\it et al.} \cite{mak};
different curves correspond to different graphene devices
(courtesy of F. Kin Mak).
For photon energies larger than 1.2 eV,
we plot data from two
different types of measurements \cite{nair}: narrow bandpass filters,
with a full band width at half maximum of 10 nm, and standard spectroscopy
measurements.
The dashed line is the result of
Eq. (\ref{eq:transmission}). The solid (green) line, to the right of 1.2 eV,
 is a plot of $T$
obtained from a calculation of the
conductivity, including trigonal warping corrections  \cite{staubergeim}.}
\label{fig:transmittance}
\end{figure}

In Fig. \ref{fig:transmittance} we show measurements from two
different groups \cite{mak,nair} in two different regions
of the spectrum. For photon energies $E_\gamma$ below 1.2 eV (in the near
infrared region), the data
were taken from reflectance measurements, at room temperature. The absorbance of
the different samples is
spectrally flat within a band of 10\%. Roughly speaking,
the transmittance follows the value given by Eq. (\ref{eq:transmission}).
A more detailed analysis  shows that
the transmittance below $E_\gamma<$0.5 eV increases over the universal value,
whereas
closer to $E_\gamma\lesssim$1.2 eV it decreases slightly from that value.
According to Mak {\it et al.} \cite{mak}, and for energies
$E_\gamma<$0.5 eV,
both temperature effects and  some amount of variable extrinsic doping
decrease  the conductivity and therefore produce an increase in the
transmittance. In other words, if light absorption decreases, then the
transmittance increases.
This effect is equivalent to a finite chemical potential (having effectively
a Pauli's principle based blocking effect), which 
contributes to an increase in the transmittance. The temperature effect
that they \cite{mak} used in their argument, was predicted to be
of importance
at room temperature and energies below $0.5$ eV  for
undoped graphene \cite{peresIJMPB}, in agreement with the experimental
measurements.
Additionally, this set of measurements showed that the optical
conductivity of graphene can also be affected by unavoidable doping
and intra-band scattering. Indeed, a SCBA \cite{nmrPRB06}
calculation of the conductivity of graphene at zero temperature
and zero doping showed that $\sigma(\omega)$ departs from its universal
value, $\sigma_0=\pi e^2/(2h)$, being strongly reduced at low frequencies, due
to
disorder.

In conclusion, the deviation of the
transmittance of graphene from the universal value predicted for
 Dirac fermions is a way of gaining insight on other
electronic effects present in the material \cite{basov}.

Figure \ref{fig:transmittance} also shows the transmittance of graphene
in the photon energy range 1.2--3 eV.
The
measured value follows the prediction of Eq. (\ref{eq:transmission}),
except at energies around 3 eV, where   absorption increases.
Since at energies as large as 3 eV the electronic energy dispersion  deviates
considerably
from the Dirac cone approximation -- an effect known as trigonal warping --
a calculation taking into account trigonal warping corrections to the 
band structure of graphene
was performed \cite{nair,staubergeim}, and the result
is given by the solid  line in  Fig. \ref{fig:transmittance}. The
calculation does predict an increase in light absorption at
energies around 3 eV (an effect opposite to that  discussed above for energies
$E_\gamma<$0.5 eV), but falls short
on accounting for the magnitude of the effect. Within the independent
electron model, the computed enhancement of the
conductivity is essentially due to the increase in the density of states
as the Van Hove singularity is approached, which is located at the 
energy of $\sim$2.7 eV. Thus, transitions between states located at the
 Van Hove singularities of the valence and the conduction bands require 
 photons of energy $\sim$5.4 eV. Optical experiments using photons of about this
energy have confirmed the strong enhancement of light absorption due to the
Van Hove singularities \cite{Geimellipsometry}.

 Two possible additional causes for such an increase in the absorption
come to mind: contamination of the sample due to some organic
residues (originated from the exfoliation process)
or/and many-body effects \cite{mishchenko,herbut}.
A recent calculation, however, showed that
electron-electron interactions correct the transmission (reducing it)
by only $0.03-0.04$\% \cite{sheehya,sheehyb,kastoptics}.
On the other hand, in the regime of energies relevant for the visible range
of the spectrum, the
fine structure constant of graphene $\alpha_g$ is a number
of order 1, a fact casting reasonable doubts on the validity of
any perturbative calculation.

A recent  {\it ab initio} calculation  \cite{LiLouie}
used the Kohn-Sham eigenvalues and eigenvectors to compute the optical
conductivity of graphene from  infrared to visible frequencies. This
type of mean-field calculations includes electron-electron interactions
by means of the exchange and correlation approximations. The 
optical conductivity obtained does not fit exactly the data of Fig.
\ref{fig:transmittance},
since it predicts an absorption higher than what
is measured experimentally, in the full frequency range. Nevertheless, the
proposed excitonic effects (included via the solution of the Bethe-Salpeter
equation) were shown to account well for the experimental deviations (to lower energies) in  
the enhancement in light absorption
for photon energies associated with
transitions between states located at the
 Van Hove singularities of the valence and the conduction bands \cite{Geimellipsometry}
relatively to the
predictions made by the 
non-interacting electrons theory.
Indeed,
the elementary theory predicts an intense absorption peak at about
5.2 eV, whereas the measured data shows a red shift of the peak to 4.6 eV. The
discrepancy
can be solved by considering the mutual attraction of the electron-hole pair,
created at the
two Van Hove singularities, when a photon of the right frequency is absorbed.

The energy red-shift mentioned is easy to understand from the point of view of
the mutual attraction of the electron-hole pair (forming an exciton): since the
electron and hole have opposite charges, the effective energy seen by the 
photon being absorbed is the non-interacting value minus a positive correction
coming from the electrostatic attraction between the electron and the hole created 
by the absorption of the photon.
%
\subsection{The optical conductivity of gated graphene}
\label{sec:sigma_gated}

Measurements of the optical conductivity of gated graphene,
in the far infrared  region of the spectrum \cite{basov}, also show
strong deviations from the simple theory given above, and expressed in concise
form by
Eq. (\ref{eq:sigoptical}).

The deviations seen in the data \cite{basov},
for all values of the gate-voltage considered in the experiment (ranging from
$V_g=$10  to 71 V),
are of five different types:
({\it i}) finite absorption below $2\mu$,
which is due to both inter-band and intra-band elastic and inelastic
scattering processes;
({\it ii}) broadening of the absorption edge
around the energy threshold $2\mu$;
({\it iii}) an enhancement of the
conductivity above the universal value $\sigma_0$ in the energy
range between $2\mu$ and $2\mu+E^\ast$, where $E^\ast$ is a
characteristic energy scale,
 ({\it iv}) a reduction of the
conductivity bellow $\sigma_0$, at energies above $E^\ast$,
with the conductivity as a function of frequency having a
positive curvature, and
({\it v}) the imaginary part of the
conductivity  is
larger than the value predicted by the non-interacting model
for energies $\hbar\omega\gg2\mu$.

Additionally, we  point out that
the optical conductivity curves measured experimentally \cite{basov}
collapse on top of each other when re-plotted as
function of $\omega/\mu$, implying that the mechanism causing
deviations from the non-interacting approximation must be intrinsic.

\begin{figure}[th]
\includegraphics*[width=8cm,angle=0]{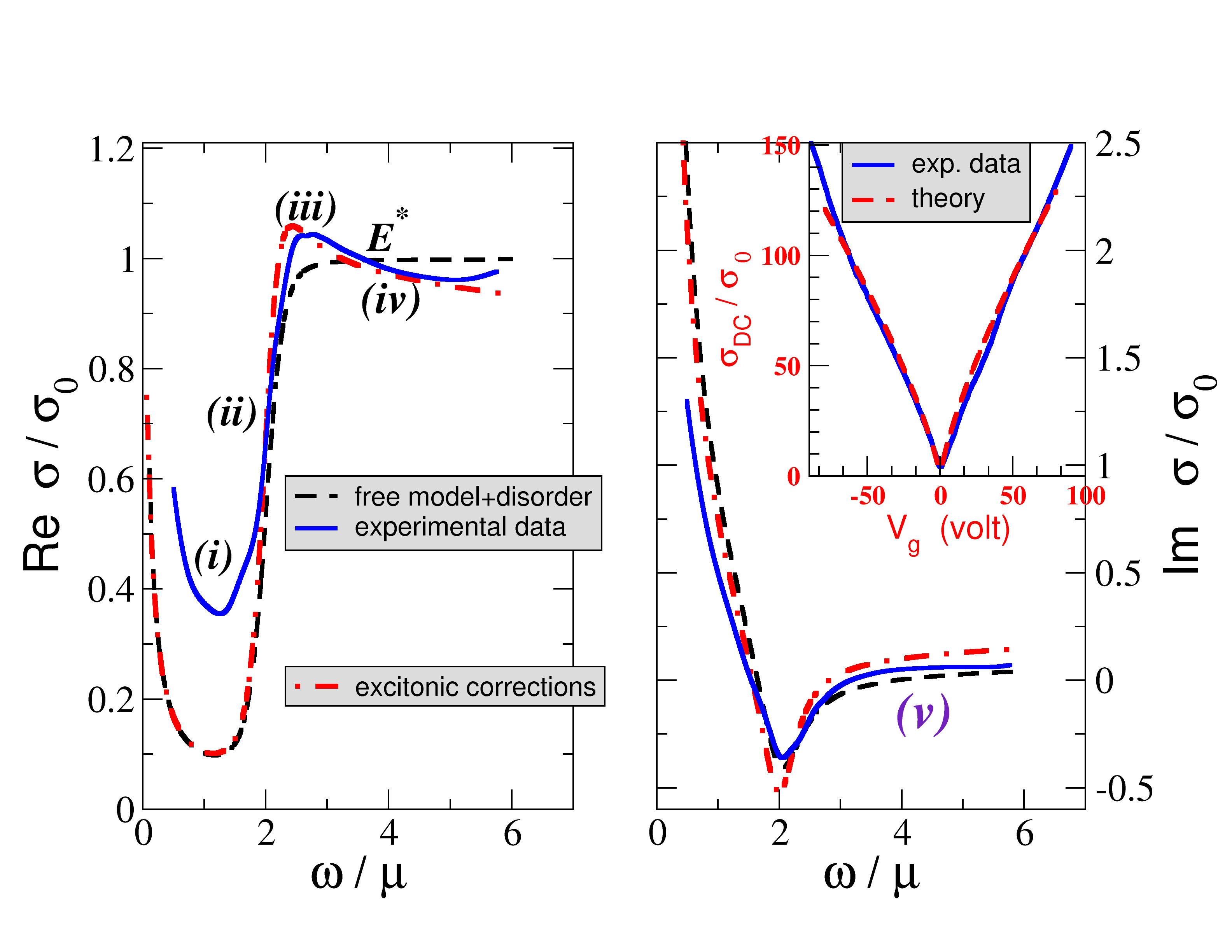}
\caption{(Color online)
Optical conductivity of gated graphene on top of silicon oxide, in units
of $\sigma_0$.
The data is given by the solid curve. The dashed  curve is a calculation taking
disorder into account, and the dot-dashed curve is a calculation including disorder
and electron-electron interactions.
The experimental curves refer to
a gate voltage of 28 V,  which corresponds a Fermi energy of $\mu\simeq0.18$
eV.
The calculation took an effective temperature of
120 K.
In the {\bf left(right)}
panel we have the real(imaginary) part of
the conductivity. In the {\bf inset},  the measured dc
conductivity of the device (solid curve) depicted
together with a fit of $\sigma(\varepsilon_F)$ using
Eqs. (\ref{eq:sigma-midgap}) and (\ref{eq:sigmaCoulomb}).
The fit of the dc conductivity fixes the concentration of impurities
to the values of $n_i=2\times 10^{11}$ cm$^{-2}$, for resonant
scatterers, and $n_i=1\times 10^{11}$ cm$^{-2}$, for charged ones (dashed curve).
(Data from Z. Li {\it et al.} \cite{basov}, courtesy of
Zhiqiang Li.)}
\label{fig:opticalgate}
\end{figure}

To explain all  measured deviations, it is necessary to take into
account disorder, temperature, electronic density inhomogeneities
\cite{Cromieinhomogeneous},
 and electron-electron interaction effects,
of excitonic nature \cite{peresexcitonic}.
In Fig. \ref{fig:opticalgate} we show one set of the measured data (solid blue
curve),
 together with calculations using two models: (i) a model where the
independent electron theory is supplemented with the effect of disorder
(computed with the
two models discussed in Sec. \ref{sec:gate}) and (ii) a calculation including
excitonic
effects on top of model (i). It is clear that the first one, considering only
disorder,
does not account for the five deviations observed in the data, relatively to the
independent
electron model; it partially accounts for the enhancement of the conductivity
below
2$\mu$. The additional optical response observed in the data, in this frequency range, 
must come from
intra-band
scattering processes \cite{juan}, not included in model (i). Points 
({\it ii})--({\it v}) are all accounted for by including excitonic
effects, that is, model (ii). There is, however,
the need to use in the model a higher temperature, $T=120$ K, than that
measured experimentally by the finger of the cryostat, $T=54$ K.
This can be justified by recalling that graphene is a system presenting
small inhomogeneities
of the electronic density. This fact effectively smears the chemical potential
and such an effect can be accounted for by considering an effective higher
temperature,
as was shown to be the case in the interpretation of 
the optical response of graphene's bilayer \cite{kuzmenkofinger}.

In the same set of measurements \cite{basov}, an increase in the
Fermi velocity over the value $1.1\times 10^6$ m/s was measured,
upon diminishing the gate voltage.
 The
 renormalized Fermi velocity,
due to exchange, was predicted to be
(including the Thomas-Fermi screening of the Coulomb potential)
\begin{equation}
v_F= v_{\rm bare}+\frac{e^2}{4\pi \epsilon_0\epsilon_{\rm d}\hbar}\left(
\ln\frac{2q_c}{9k_F}-\frac{1}{3}\right)\,,
\label{eq:VFrenorm}
\end{equation}
where $v_{\rm bare}$ is the bare Fermi velocity, and therefore not an observable
parameter, $q_c\sim 1$ \AA$^{-1}$ a cutoff momentum, and we have used the
fact that the Thomas-Fermi screening momentum is for graphene
on silicon oxide given by $q_{\rm TF}\simeq 2k_F$ [we note that Eq. (\ref{eq:VFrenorm})
is an extension, taking screening into
account, of previous results \cite{MedianoVF,sarnexchange,poliniB}].
The result (\ref{eq:VFrenorm}) does show that $v_F$ increases upon decreasing
$k_F$, but the formula fails to fit the data \cite{basov}
over the measured gate voltage range,
especially for larger $V_g$.
Alternatively,
phonons also seem to partially account for the Fermi velocity renormalization
seen in
the experiments \cite{carbotte,stauberphonons,peresEPL}.

The renormalization of the Fermi velocity  suggests
that contributions from many-body effects  can be important, but
an experiment with a higher degree of accuracy should be repeated, since
cyclotron mass measurements \cite{qhegeim} do not see deviations of the Fermi
velocity from $v_F\simeq 1.1\times 10^6$ m/s upon varying the electronic
density.

Finally, we note that, as in the case of neutral graphene discussed
in Sec. \ref{sec:sigma_neutral}, the electron-hole pair mutual-attraction
also induces a shift in the Fermi edge at $\omega=2\mu$ toward lower
energies (red shift effect), since, as before, it renormalizes the non-interacting
energies by adding a negative number. 

In conclusion, the optical response of graphene is a property of the material
where different kinds of quantum effects seem to play an important role, all on
equal footing. Exploring graphene for nanophotonic devices
surely requires a detailed understanding of its optical properties.

\section{Conclusions}

We have given an introductory review on the  transport properties of graphene,
touching
 the relevant aspects of this topic. A survey of the literature was
given along with a discussion of elementary models, which,
despite their simplicity, are in
good quantitative agreement with many features of the data.
Many of these models were discussed in
greater detail in the literature (to which due credit is given),
but using more formal methods. We hope that the topics discussed in this
Colloquium
may contribute to a wider dissemination
of the physics of graphene among the non-specialized audience.

\begin{acknowledgments}
I have benefited immensely from discussions with
many colleagues and friends in the last few years, but I would like
to thank especially, Aires Ferreira, Andre Geim,
Antonio Castro Neto, Bruno Uchoa, Eduardo Castro,
Eduardo Mucciolo,
Fernando Sols,
Francisco Guinea, Jaime Santos, Jo\~ao Lopes dos Santos,
Johan Nilsson, Jos\'e Carlos Gomes,
Konstantin Novoselov,  Maria Vozmediano,
Mikhail Vasilevskiy,
 Ricardo Ribeiro, Shan-Wen Tsai, Tobias Stauber,
Vitor Pereira, and Yuliy Bludov.
I also would like to thank the anonymous referees since their
comments were most useful to improve the final version of the paper.
The hospitality of the National University of Singapore (NUS) is
acknowledge, where this work was started.
 A special thanks
goes to ALCRP, FCRP, and MAIC, for their continuous support.
\end{acknowledgments}

\bibliographystyle{apsrmp4-1} 

%

\end{document}